\documentclass[%
reprint, 
 amsmath,amssymb,
 aps, 
pre,
]{revtex4-2}

\usepackage{graphicx}% Include figure files
\usepackage{dcolumn}% Align table columns on decimal point
\usepackage{bm}% bold math
\usepackage[dvipsnames]{xcolor}
\begin{document}

% \preprint{APS/123-QED}

\title{Ordering-disordering dynamics of the $q$-voter model under random external bias}

\author{Roni Muslim\textsuperscript{1,2}}
\email{roni.muslim@apctp.org}
\email{roni.muslim@brin.go.id}

\author{Jihye Kim\textsuperscript{3}}
\email{arrr3755@naver.com}

\author{Noriko Oikawa\textsuperscript{4}}
\email{noriko.oikawa@omu.ac.jp}

\author{Rinto Anugraha NQZ\textsuperscript{5}}
\email{rinto@ugm.ac.id}

\author{Zulkaida Akbar\textsuperscript{2}}
\email{zulkaida.akbar@brin.go.id}

\affiliation{\textsuperscript{1}Asia Pacific Center for Theoretical Physics, POSTECH, Pohang, 37673, Republic of Korea}
\affiliation{\textsuperscript{2} Research Center for Quantum Physics, BRIN, South Tangerang, 15314, Indonesia}
\affiliation{\textsuperscript{3}Department of Physics, Korea University, Seoul, 02841, Republic of Korea}
\affiliation{\textsuperscript{4}Department of Physics and Electronics, Osaka Metropolitan University, Sakai, 599-8531, Japan}
\affiliation{\textsuperscript{5}Department of Physics, Universitas Gadjah Mada, Yogyakarta, 55282, Indonesia}

\date{\today}% It is always \today, today,

\begin{abstract}
We investigate a variant of the two-state $q$-voter model in which agents update their states under a random external field (which points upward with probability $s$ and downward with probability $1-s$) with probability $p$ or adopt the unanimous opinion of $q$ randomly selected neighbors with probability $ 1-p$.  Using mean-field analysis and Monte Carlo simulations, we identify an order-disorder transition at $p_c$ when $s=\tfrac{1}{2}$.  Notably, in the regime of $p>p_c$, we estimate the time for systems to reach disordered state from consensus state and find the logarithmic scaling $T_{\text{dis}} \sim \mathcal{B}\ln N$, with $\mathcal{B} = 1/(2p)$ for $q = 1$, while for $q > 1$, $\mathcal{B}$ depends on both $p > p_c$ and $q$.  We observe that disordering dynamics slow down significantly for nonlinear strengths $q$ between $2$ and $3$, independent of the probability $p$.  On the other hand, when $s=0$ or $s=1$, the system is bound to reach consensus, with the consensus time scaling logarithmically with system size as $T_{\text{con}} \sim \mathcal{B}\ln N$, where $\mathcal{B} = 1/p$ for $q = 1$ and $\mathcal{B} = 1$ for $q > 1$. Furthermore, in the limit of $p = 0$, we derive a closed-form exit probability valid for arbitrary values of $q$ and demonstrate a finite-size scaling collapse.  These results clarify how external cues and peer conformity jointly control ordering and disordering in binary opinion dynamics.

% These results provide insights into how bipolar media environment and peer pressure jointly govern the opinion dynamics in social systems.
\end{abstract}

\maketitle

% \tableofcontents
\section{\label{sec:Sec1}Introduction}

The voter model (VM), in which an agent $i$ holds one of two possible states (i.e., opinions) $\sigma_i=1$ or $\sigma_i=-1$, has long served as a paradigmatic framework for modeling binary opinion dynamics in statistical physics~\cite{liggett1999stochastic, castellano2009statistical, redner2019reality}. Over the past two decades, extensive efforts~\cite{suchecki2005voter, sood2005voter, peralta2020ordering, ramirez2022local, muslim2024mass, kim2024competition,  anugraha2025nonlinear, schweitzer2018sociophysics} have been made to generalize the VM and its nonlinear variants to account for more realistic social and physical features, including heterogeneous behaviors~\cite{nyczka2012phase, mobilia2015nonlinear, mellor2017heterogeneous}, external influences~\cite{civitarese2021external, muslim2024mass, doniec2025modeling, fardela2025opinion, azhari2023external}, and complex interaction structures such as small-world, scale-free, and hypergraph topologies~\cite{jkedrzejewski2017pair, vieira2018threshold, vieira2020pair, meyer2024time, kim2024competition, fardela2025opinion, anugraha2025nonlinear, sood2008voter}. These models encapsulate the essential mechanisms of consensus formation, disagreement, and stochastic fluctuations in populations of interacting agents. In particular, the $q$-voter model ($q$-VM), the best-known nonlinear variant of the VM, introduces a local interaction rule in which an individual adopts the unanimous opinion of a randomly selected group of $q$ neighbors (the $q$-panel). By contrast, when the $q$-panel is not unanimous, individual opinions flip at random with probability $\varepsilon$ \cite{castellano2009nonlinear}.

In studies of VMs, the ordering dynamics of the system of size $N$ is characterized by  the order parameter $m\equiv\sum_{i} \sigma_{i}/N$. The system might transition between $m=0$ (i.e., `disordered') and $m\neq0$ (i.e., `ordered') states. A key quantity in VMs is the exit time (or consensus time), the average time required for the system to reach an absorbing consensus (i.e., $m\pm 1$) state. It has been well established that the consensus time depends strongly on the system size $N$, the interaction topology, and the specific update rules. For instance, in the standard VM on complete graphs and Erdős–Rényi networks, the consensus time scales linearly with system size~\cite{castellano2005comparison, sood2005voter}, while on Barabási–Albert networks, it grows sublinearly as $T \sim N/\ln N$. In contrast, for nonlinear extensions such as the $q$-VM with $q > 1$, the consensus time has been shown to scale logarithmically or even exponentially with $N$, depending on the model parameters and update mechanisms~\cite{meyer2024time, ramirez2024ordering, kim2024competition}.

Another quantity of central interest is the exit probability, defined as the likelihood that the system reaches a particular absorbing state as a function of the initial condition. In classical VMs, the exit probability exhibits a linear dependence on the initial fraction of opinions~\cite{sood2005voter}, while in nonlinear models, the behavior becomes highly nontrivial~\cite{mobilia2003does, peralta2020ordering}. Analytical and numerical investigations have shown that the shape of the exit probability curve encodes key information about underlying symmetries, update rules, and the strength of nonlinearity~\cite{muslim2024mass, doniec2025modeling}. Recent work showed that introducing a weighted‐influence mechanism in the $q$-voter model, where an influence‐power parameter modulates the adoption probability of the majority opinion, profoundly affects both the consensus‐time scaling and the form of the exit‐probability function~\cite{mullick2025social}.

Within the $q$-voter framework, the interplay between conformity, namely, agents adopt the unanimous opinion of a randomly selected panel of $q$ neighbors, and independence, namely, agents update their opinion independently of their neighbors, has been shown to induce both continuous and discontinuous phase transitions, depending on the values of the nonlinear strength $q$~\cite{nyczka2012phase, chmiel2015phase, jkedrzejewski2017pair, anugraha2025nonlinear}. These transitions mark the shift between disordered and ordered states. Analytical approaches based on mean-field theory and pair approximations have provided insights into the nature of phase boundaries~\cite{jkedrzejewski2017pair}. Moreover, recent studies have examined relaxation dynamics and domain coarsening, revealing that nonlinearity significantly alter the ordering kinetics~\cite{corberi2024ordering, corberi2024coarsening}.

In this work, we investigate a variant of the $q$-voter model in the presence of a random external field. At each update, an agent either acts independently with probability $p$, adopting the favored opinion, $+1$ with probability $s$, and $-1$ with probability $1-s$, or with probability $1-p$ conforms to the unanimous opinion of a randomly selected $q$-panel. The key difference between the original $q$-VM and the present formulation concerns how the independence or noise mechanism is implemented. In the original model~\cite{castellano2009nonlinear}, the parameter $\varepsilon$ determines the probability of flipping opinions when the $q$-panel is not unanimous. In contrast, in the present model, the independence probability $p$ applies regardless of the $q$-panel's configuration.

The present model provides a minimal yet flexible setting to capture the influence of external propaganda on collective opinion dynamics. We analyze the model using mean-field theory and corroborate the analytical predictions with Monte Carlo (MC) simulations. We demonstrate that an asymmetric (symmetric) bias $s\neq \tfrac{1}{2}$ ($s= \tfrac{1}{2}$) drives a system towards an ordered (disordered) state. In particular,
when $s=0$ or $s=1$, the system reaches consensus as in the case of $p=0$. By contrast, when $s=\tfrac{1}{2}$ and $p$ exceeds a critical threshold, the system invariably relaxes to a vanishing magnetization ($m=0$) regardless of its initial condition. To quantify this relaxation process, we introduce the mean disordering time (hereafter, ``disordering time"), defined as the characteristic time required for the system to transition from consensus to a disordered state. The disordering time serves as a central metric in $q$-voter models, offering comprehensive insight into the ordering–disordering dynamics that have not been addressed in previous studies.

Our focus is on analyzing how consensus time, disordering time, and exit probability depend on system size, nonlinear strength $q$. For all nonlinear strengths $q > 1$, the consensus time exhibits a universal logarithmic scaling with system size $N$. In contrast, the linear case $q = 1$ scales inversely with independence probability $p$, consistent with memoryless dynamics. Additionally, we find that the disordering time scales logarithmically with system size, with a prefactor dependent on the independence probability $p$ and nonlinear strength $q$. Notably, disordering time dynamics slow down significantly for nonlinear strengths $q$ between $2$ and $3$, independent of $p$. Furthermore, we derive an explicit analytical expression for the exit probability, analyze its scaling behavior, and verify its accuracy via numerical simulations. We also identify a saturation regime in the large-$q$ limit, where nonlinear effects become negligible.

These findings provide new insights into how the interplay between global bias and local consensus mechanisms shapes opinion dynamics, thereby contributing to a deeper understanding of collective behavior in opinion formation models.

%%%%%%%%%%%%%%%%%%%%%%%%%%%%%%%%%%%%%%%%%%%%%%%%%%%%%%%%%%%%%%%%%%%%%%%%%%%%%%%%
\section{Model Description}
\label{sec.2}
\begin{figure}[tb]
\centering
\includegraphics[width = 0.8\linewidth]{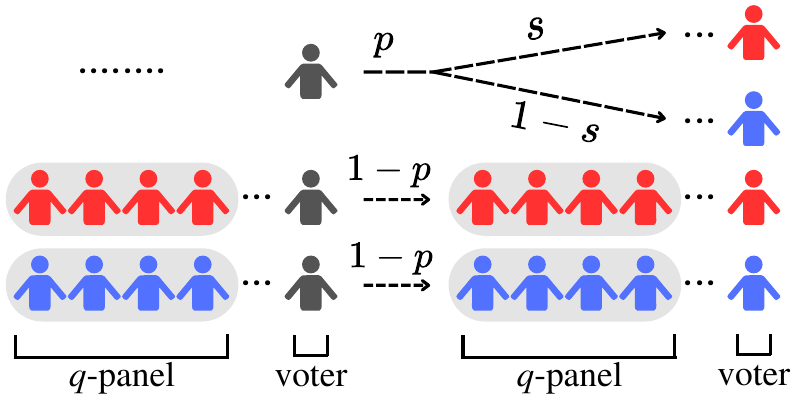}
\caption{Illustration of the $q$-VM under a random external field. Red and blue represent opinions $+1$ and $ -1$, respectively. At each update, a voter (black agent) is randomly selected. With probability $p$, the voter acts independently of its nearest neighbors, adopting opinion $+1$ with probability $s$ or $-1$ with probability $1-s$. Conversely, with probability $1-p$, it consults a randomly selected panel of $q$ neighbors (the $q$-panel) and adopts their opinion only if the panel is unanimous.}
\label{fig:illustra}
\end{figure}

We study the $q$-voter model with independence in the sense of Refs.~\cite{nyczka2012phase, jkedrzejewski2017pair}.  At each update, an individual or a voter and a group of $q$ neighbors (the $q$-panel) are sampled at random.  With probability $p$, the voter acts independently, choosing state $+1$ with probability $s$ and $-1$ with probability $1-s$, irrespective of the $q$-panel’s composition as illustrated in Fig.~\ref{fig:illustra}.  With probability $1-p$, the voter conforms if the $q$-panel is unanimous; otherwise, the voter’s state remains unchanged.  For $s=\tfrac{1}{2}$, this reduces to the symmetric independent-flipping variant analyzed in Refs.~\cite{nyczka2012phase, jkedrzejewski2017pair}.

We consider an undirected network of $N$ agents who reside in the nodes and interact with their nearest neighbors. The edges of the network represent social connections. In this work, we focus on the mean-field limit by considering a complete graph, where each agent interacts with all others (all-to-all interaction). 

At each time step $t$, the fraction $c(t)$ of agents with the opinion $+1$ evolves by the following process:
$(1)$ an agent and its $q$ neighbors (i.e., $q$-panel) are picked uniformly at random. $(2)$ With probability $p$, the agent behaves independently of the $q$-panel by adopting $+1$ with probability $s$, or $-1$ with probability $1 - s$. (3) With the remaining probability $1 - p$, the agent adopts the unanimous state of the $q$-panel, if unanimity is present. Otherwise, the agent remains unchanged. It is worth noting that there is a spin-flip symmetry: replacing $s$ by $1-s$ together with relabeling $+1\leftrightarrow-1$ maps $c$ to $1-c$ (and $c(0)$ to $1-c(0)$); hence it suffices to analyze $s=1$ since the case $s=0$ follows by symmetry.

In our formulation, the independence $p$ is the frequency individuals ignore local peer composition and update from an exogenous cue; the bias $s$ measures that cue’s directional strength toward $+1$ (and $1-s$ toward $-1$). When $s=\tfrac{1}{2}$, the cue is unbiased and behaves as symmetric noise. For $s\neq\tfrac{1}{2}$, it induces a drift analogous to an external field in the Ising model~\cite{brush1967history}. Socially, $p$ captures reliance on institutional guidance, mass media, or stable priors under mixed local signals~\cite{cohen2003party,druckman2013elite}, whereas the conformity branch activates only under strong local consensus (unanimous $q$-panel), echoing unanimity effects and social-impact theory~\cite{latane1981psychology}. If the panel is non-unanimous, the agent maintains the status quo. Typical instances include organizational adherence, consumer adoption, and political choice when external cues substitute for local consensus~\cite{chevalier2006effect, bavel2020using}. The unanimity requirement aligns with threshold and complex-contagion mechanisms requiring multiple simultaneous reinforcements~\cite{centola2007complex}.

This formulation allows us to investigate how the interplay between group conformity, individual independence, and external fields shapes the collective behavior of the system. Two external fields, one favoring the opinion $+1$ with the relative strength $s$ and another favoring the opinion $ -1$ with the relative strength $1-s$, can prevent the system from reaching consensus. If one external field exists, i.e., $s = 0$ or $s = 1$, the system always reaches a fully ordered state, with all agents sharing the same opinion.

\section{Time Evolution and Stationary}
\subsection{Time Evolution}
In the mean-field theory, the macroscopic state of the system is fully characterized by the fraction $c$ of agents in the $+1$ state. Equivalently, $c$ can be interpreted as the probability of finding an individual in state $+1$, or related to the system's order parameter via $c = (m + 1)/2 \in [0, 1]$.

A single agent is selected and updated at each elementary time step $\delta t = 1/N$. The system may undergo one of three transitions: a flip from $+1$ to $-1$, a flip from $-1$ to $+1$, or no change. Each update results in a change of $c$ by an increment $\delta c = 1/N$. The probability of increasing (raising) or decreasing (lowering) $c$ by $\delta c$ is denoted by $R(c)$ and $L(c)$, respectively.

Under the mean-field approximation, the raising and lowering probabilities in the thermodynamics limit are given by [see Appendix~\ref{appendix_A}]:
\begin{align}
    R(c) & = (1 - c) \left[(1 - p)\,c^q + ps\right], \label{eq:vara_12} \\
    L(c) & = c \left[(1 - p)\,(1 - c)^q + p(1 - s)\right]. \label{eq:varb_13}
\end{align}
The time evolution of $c$ is governed by the recurrence relation $c(t + \delta t) = c(t) + [R(c) - L(c)]/N$~\cite{krapivsky2010kinetic}, which, in the continuous-time limit, becomes
\begin{equation}\label{eq:time_contin} 
    \int_{c(0)}^{c(t)} \frac{dc'}{R(c') - L(c')} = t.
\end{equation}
Since an explicit solution for $c(t)$ is generally intractable for arbitrary values of $q$, $p$, and $s$, one may instead consider the implicit analytical solution to Eq.~\eqref{eq:time_contin}, which reveals key features of the system’s long-time behavior.

The implicit solution can be expressed as:
\begin{equation}\label{eq:implicit}
    \prod_{i=1}^{n} \left| \frac{c(t) - r_i}{c(0) - r_i} \right|^{\dfrac{1}{\prod_{j \neq i} (r_i - r_j)}} = \exp\left[{-K(q, p)\, t}\right],
\end{equation}
where $r_i$ $(i = 1, \dots, n)$ are the real or complex roots of the drift equation $v(c) = R(c) - L(c)$, and $n$ is the degree of the polynomial $v(c)$, determined by the nonlinearity $q$ as follows [see Appendix~\ref{appendix_B}]:
\begin{equation}
    n =
\begin{cases}
1, & q = 1, \\
q + 1, & q > 1 \quad \text{and $q$ even}, \\
q, & q > 1 \quad \text{and $q$ odd}.
\end{cases}
\end{equation}
The coefficient $K(q, p)$ sets the overall time scale of the exponential relaxation and is given by
\begin{equation}
    K(q, p) =
\begin{cases}
p, & q = 1, \\
2(1 - p), & q > 1 \quad \text{and $q$ even}, \\
(q - 1)(1 - p), & q > 1 \quad \text{and $q$ odd}.
\end{cases}
\end{equation}

In the asymptotic limit $t \to \infty$, the right-hand side of Eq.~\eqref{eq:implicit} vanishes, implying that $c(t)$ converges to one of the stable fixed points, denoted $r_s$, which satisfies $v(r_s) = 0$ and $v'(r_s) < 0$. A linear stability analysis near $r_s$ shows that small deviations from equilibrium decay exponentially, leading to the approximation $c(t) \approx r_s + \epsilon(0) \exp[v'(r_s)t]$, where $\epsilon(0)$ is the initial deviation.

For example, in the linear case, where $q = 1$, the drift function $v(c)$ is linear, and the unique fixed point is $r_1 = s$, which is stable. Substituting $r_1 = s$ into Eq.~\eqref{eq:implicit}, one obtains an explicit solution for $c(t)$ as:
\begin{equation}\label{eq:linear_c}
    c(t) = c(0)\, \exp(-p t) + s\left[1 - \exp(-p t)\right].
\end{equation}
This expression describes exponential convergence toward the equilibrium value $s$, with a characteristic relaxation time that scales as $\sim 1/p$. For $p = 0$, the drift term vanishes identically, and $c(t)$ remains constant in time: $c(t) = c(0)$. This recovers the well-known result for the classical linear VM~\cite{clifford1973model, liggett1999stochastic}, in which the dynamics are purely diffusive and exhibit no directional bias without external fields.

For $q = 2$ and $q = 3$, the drift function $v(c)$ becomes cubic, and its fixed points can be obtained analytically via Cardano’s method~\cite{witula2010cardano}. The three roots $r_i$ ($i = 0, 1, 2$) are given by:
\begin{align}\label{eq:r_q2}
    r_i = \frac{1}{2} &+ \sqrt{\frac{1 - 3p}{3\left(1 - p\right)}}  
    \cos\left[\frac{1}{3}\arccos\theta- \frac{2\pi(i - 1)}{3}\right],
\end{align}
where $\theta = [3p\left(1 - 2s\right)\sqrt{3\left(1 - p\right)}]/[\left(1 - 3p\right)^{\tfrac32}]$, and the roots are real for $p \leq 1/3$. For $p > 1/3$, only one real root exists, and the remaining two form a complex conjugate pair, indicating a qualitative change in the fixed-point structure of the system. Unlike the linear case, it is generally impossible to write a closed-form expression for $c(t)$ when $q > 1$. Nevertheless, the time evolution can in principle be obtained by substituting the roots $r_i$ into Eq.~\eqref{eq:implicit}.

For $p = 0$, Eq.~\eqref{eq:r_q2} yields three real roots: $r_0 = 0$, $r_1 = 1$, and $r_2 = 1/2$. The long-time behavior of $c(t)$ is governed by these fixed points, and the particular attractor approached depends solely on the initial condition. In this case, an explicit expression for the time evolution can be written as:
\begin{equation}
    c(t) = \frac{1}{2} \left[ 1 + \frac{\left[2\,c(0)-1\right]\exp[\frac{t}{2}]}{\sqrt{4\,c(0)\left[1-c(0)\right] + \left[2\,c(0)-1\right]^2 \exp(t)}} \right],
\end{equation}
which describes deterministic convergence toward either $r_0 = 0$ or $r_1 = 1$. The direction of convergence is determined entirely by the initial condition: the system evolves toward $r_0 = 0$ if $c(0) < \tfrac{1}{2}$, and toward $r_1 = 1$ if $c(0) > \frac{1}{2}$.

\begin{figure}[tb]
    \centering
    \includegraphics[width = \linewidth]{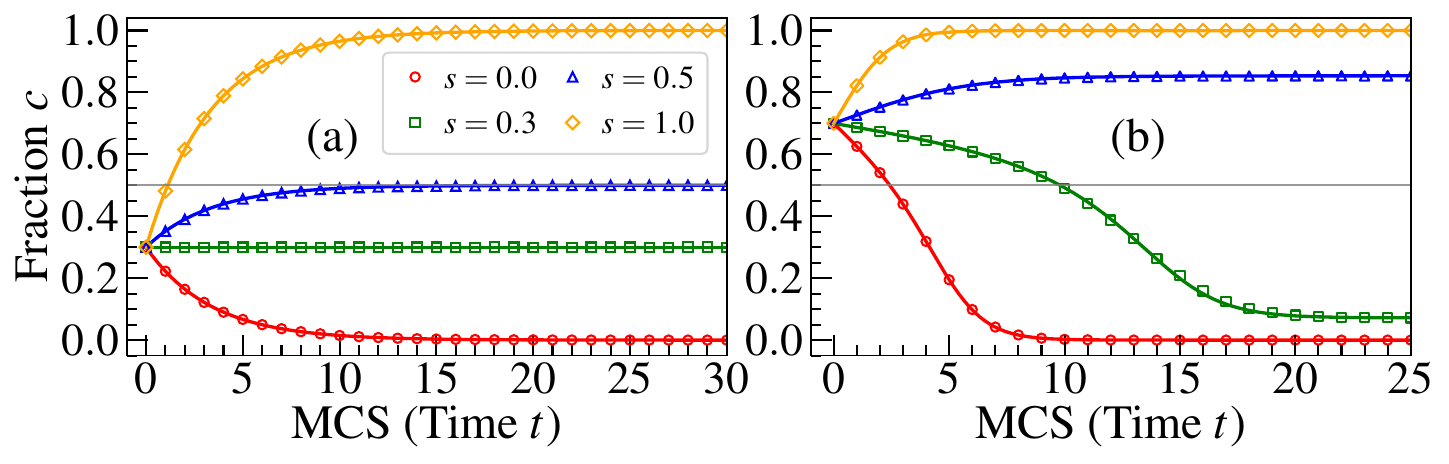}
    \caption{Time evolution of the state fraction $c(t)$ for various values of the bias parameter $s$. Symbols denote MC simulation results, while solid lines represent analytical predictions. Panel (a): results for $q = 1$, with parameters $p = 0.3$, $c(0) = 0.3$. Panel (b): results for $q = 2$, with $p = 0.2$ and $c(0) = 0.7$. The fixed points in this case are $c \approx 0.0718$ for $s = 0.3$ and $c \approx 0.8533$ for $s = 0.5$. In both panels, the system size is $N = 10^4$, and data are averaged over $10^4$ realizations.}
    \label{fig:mcs}
\end{figure}

To validate these analytical predictions, we compare them with MC simulations in Fig.~\ref{fig:mcs}. The simulation results (symbols) agree with the theoretical curves (solid lines). The results indicate that consensus, i.e., a fully ordered absorbing state, is achieved only in the limiting cases $s = 0$ and $s = 1$.  In contrast, for intermediate values $0 < s < 1$, the system fails to reach consensus due to the competition between the opposing external biases, which becomes particularly evident in the case of $q=1$ and $p>0$, where the stationary-state value of $c$ equals $s$. 

The behavior of fixed points of the drift function $v(c)$, concerning $q$, $p$, and $s$, is so complicated that determining the exact stability landscape is generally intractable. However, qualitative features of the dynamics can still be inferred by analyzing the fixed points and their local stability. Fixed points are defined by the condition $v(c) = 0$, and their stability is determined by the sign of the derivative $v'(c)$: a fixed point is stable if $v'(c) < 0$ and unstable if $v'(c) > 0$.

To gain insight, we examine three representative cases: $s = 0$, $s = 1$, and $s=\tfrac{1}{2}$. For $s = 0$, the system is biased toward the $-1$ state and the stable fixed point is always $c= 0$. For $s = 1$, the system is biased toward the $+1$ state, and the stable fixed point is $c= 1$. The system is symmetric for $s=\tfrac{1}{2}$, and the behavior depends critically on the value of $p$. For $s=\tfrac{1}{2}$, the drift function admits a central fixed point at $c=1/2$ whose linear stability changes sign at the critical probability $p_c$: it is stable for $p>p_c$ (disordered phase) and unstable for $0<p<p_c$ (ordered phase).  In the ordered regime, spontaneous symmetry breaking drives the dynamics away from $c=1/2$, and the stationary condition of the drift function yields two interior stable fixed points $c_{\pm}=1/2\pm\delta$, where [see Appendix~\ref{appendix_C}]
\begin{equation}
    \delta \approx \left[\frac{(1 - p)(q - 1) 2^{1 - q} - p}{B_{2k+1}(q)}\right]^{1/2k},
\end{equation}
with $B_{2k+1}(q)$ being the first nonzero coefficient in the odd-order expansion of the drift function $v(c)$ around $c = 1/2$, and $k$ denotes the smallest positive integer such that $B_{2k+1}(q) \neq 0$.

As an illustrative, for $q = 2$, the dominant nonlinear contribution arises at cubic order ($k = 1$), with $B_{3}(2) = 4(1 - p)(q - 1) 2^{1 - q}$. Substituting into the expression for $\delta$, we obtain $\delta \approx \sqrt{(1-3p)/4(1-p)}$ which holds for $p < 1/3$. For instance, at $p = 0.2$, this yields $\delta \approx 0.3536$, resulting in two symmetric stable fixed points located at $c_- \approx 0.1464$ and $c_+ \approx 0.8536$, as shown in panel (b) of Fig.~\ref{fig:mcs} for the case $s = 0.5$.

\subsection{Stationary State and Phase Transition}

To characterize the behavior of the system, we examine the stationary-state value of $c$ when the drift function becomes zero. The independence probability $p$ can be expressed as:
\begin{equation}\label{eq:stat}
    p(c, q, s) = 
    \frac{c^{1+q} + c \left(1 - c\right)^q - c^q}
         {c^{1+q} + c \left(1 - c\right)^q - c^q - c + s}.
\end{equation}
This relation shows that the stationary independence probability $p$ is modulated by the external bias $s$. Although Eq.~\eqref{eq:stat} cannot be inverted analytically for arbitrary $q$, it can be evaluated numerically to obtain $c$ for given $p$ and $s$.

Nevertheless, Eq.~\eqref{eq:stat} enables efficient computation of $c(p)$ curves at fixed $s$, facilitating analysis of the system's ordering behavior. At $s = \tfrac{1}{2}$, a phase transition occurs when the system crosses the symmetry point $c = 1/2$, corresponding to a vanishing order parameter $m = 0$. The critical point $p_c$ for this transition can be obtained by evaluating Eq.~\eqref{eq:stat} at $\lim{c \to 1/2} $. For the symmetric case $s=\tfrac{1}{2}$, this yields $p_c = (q - 1)/(q - 1 + 2^{q - 1})$, which coincides with the mean-field critical point of the $q$-VM with independence~\cite{nyczka2012phase, jkedrzejewski2017pair, abramiuk2019independence, civitarese2021external, anugraha2025nonlinear}.

For the asymmetric case $s\neq\tfrac{1}{2}$, the up–down symmetry is explicitly broken and the pitchfork at $s=\tfrac{1}{2}$ unfolds. When $1<q<5$ (supercritical regime), the unfolding produces a pair of saddle-node bifurcations delimiting a finite bistable interval (a cusp catastrophe). When $q>5$ (subcritical regime), the symmetric subcritical pitchfork unfolds into a three-to-one sequence of saddle–node bifurcations (a butterfly catastrophe). Thus, for small asymmetry $\delta s\equiv |s-\tfrac{1}{2}|\ll 1$, the bifurcation structure persists with multistability, whereas sufficiently large $\delta s$ leaves a single stable state. This scenario mirrors the nonlinear noisy voter model analyzed in Ref.~\cite{peralta2018analytical}, where a tricritical point separates the supercritical and subcritical regimes, and asymmetry generates cusp (for $1<q<5$) and butterfly (for $q>5$) lines.

\begin{figure}[tb]
    \centering
    \includegraphics[width=\linewidth]{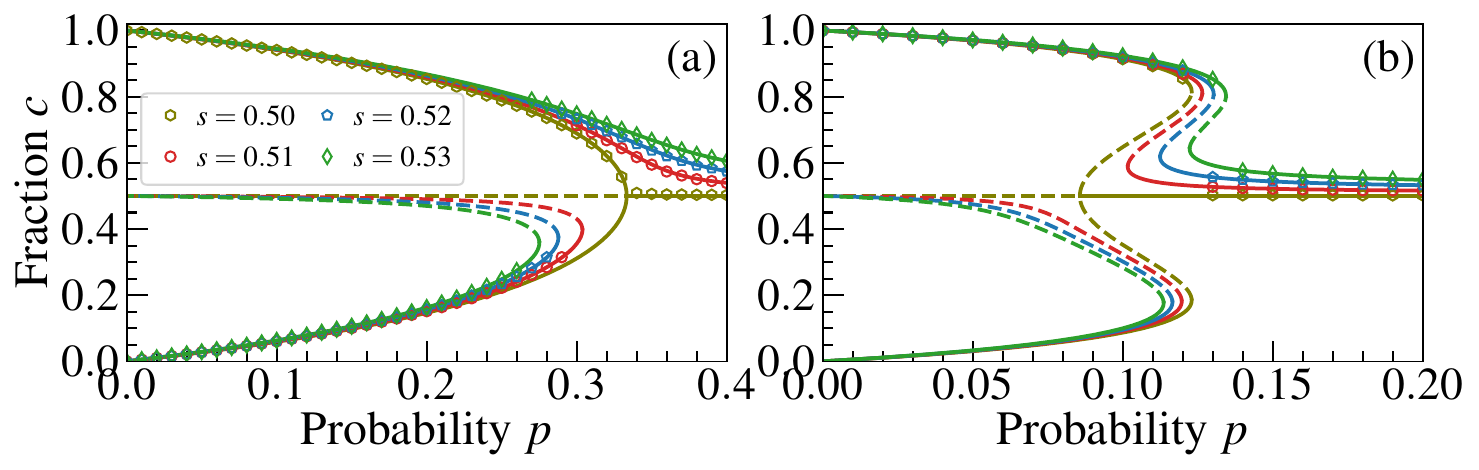}
     \caption{Stationary fraction $c$ versus independence probability $p$ for $s\in\{0.50,0.51,0.52,0.53\}$. Solid lines show stable mean-field fixed points from Eq.~\eqref{eq:stat} and dashed lines show unstable branches. Symbols denote MC results on a fully connected graph with $N=10^4$, each point averages $10^5$ realizations, and the initial condition of the simulation is homogeneous $+1$. (a) $q=3$ (supercritical): at $s = \tfrac{1}{2}$ a pitchfork occurs. For $s > \tfrac{1}{2}$ it unfolds into an imperfect (cusp-like) bifurcation with finite bistability. (b) $q=7$ (subcritical): at $s=\tfrac{1}{2}$ the transition is subcritical with metastable branches, a small asymmetry $s> \tfrac{1}{2}$ unfolds it into butterfly-type saddle-node sequences, yielding multistability that shrinks as $|s-\tfrac{1}{2}|$ increases.}
    \label{fig:phase_var_s}
\end{figure}

Figure~\ref{fig:phase_var_s} shows the stationary fraction $c$ as a function of the independence probability $p$ for several values of $s$. Solid (dashed) curves are stable (unstable) fixed points from Eq.~\eqref{eq:stat}, and symbols are MC data on a fully connected graph initialized from a homogeneous $+1$ state. The analytical predictions and simulations agree across all $s$. At the symmetric point $s=\tfrac{1}{2}$, the transition is continuous for $q=3$ (a supercritical pitchfork) and discontinuous for $q = 7$ (a subcritical pitchfork). Near $p_c$ at $s=\tfrac12$ with $1<q<5$, the deviation from symmetry scales as $|c-1/2|\sim(p-p_c)^{1/2}$ (equivalently $m=2c-1\sim(p-p_c)^{1/2}$), consistent with mean-field Ising criticality. As discussed above, for $s\neq\tfrac{1}{2}$ the unfolded cusp and butterfly structure yields finite intervals of bistability that shrink as $\delta s$ increases. A global view of these bifurcations in the $(s,p)$ plane is obtained by plotting the fold (saddle–node) loci (see Appendix~\ref{appendix_C}, Fig.~\ref{fig:phase_ps}). 
The multistable domain in that diagram corresponds precisely to the $p$–intervals where multiple fixed points appear in Fig.~\ref{fig:phase_var_s}.

\section{Consensus Time}
Two key dynamical observables in the voter models are the mean consensus time (hereafter, ``consensus time”) and the exit probability. The consensus time is the mean first-passage time to an absorbing consensus state (all-$+1$ or all-$-1$), given an initial fraction $c(0)$, under absorbing boundaries at $c=0$ and $c=1$. All agents share the same opinion in that state, and no further changes occur.
% Two key dynamical observables in VMs are the mean consensus time (hereafter, ``consensus time") and the exit probability. The consensus time quantifies the mean first-passage time to an absorbing consensus state (all-$+1$ or all-$-1$), starting from an initial condition $c(0)$. In that state all agents share the same opinion and no further changes occur.

In the model considered here, consensus can be achieved under two distinct limiting conditions: (1) Absence of external fields: When the independence probability vanishes, i.e., $p = 0$, agent updates occur solely via interaction with a unanimous $q$-panel. In this case, the dynamics are entirely deterministic in the thermodynamic limit, and the final state depends only on the initial condition. If $c(0) > 1/2$, the system evolves to $c = 1$, while for $c(0) < 1/2$, it evolves to $c = 0$. (2) Absence of competition between external fields. When the bias parameter is fixed at either $s = 0$ or $s = 1$, the external field deterministically drives the system to the $-1$ or $+1$ absorbing state, respectively. In this regime, the consensus time decreases monotonically with increasing $p$.

In general, the consensus time is influenced by the drift $v(c) = R(c) - L(c)$ and the diffusion function $D(c) = [R(c) + L(c)]/(2N)$. While obtaining a closed-form solution for $T(c)$ is generally intractable due to the nonlinear structure of $v(c)$ and $D(c)$, well-performing approximate solutions can be obtained in the large-$N$ limit with negligible diffusion. Within this approximation, the consensus time from the initial fraction (condition) $c(0)$ is
\begin{equation}\label{eq:exit_approx}
T(c(0)) \approx \int_{c(0)}^{1-1/N} \frac{dc}{v(c)},
\end{equation}
where the upper limit reflects the absorbing boundary at $c=1$. For $s=1$ and $p >0$, $c=1$ is absorbing; since $c$ takes values on the lattice $\{0, 1/N, \ldots, 1\}$, the nearest interior state is $1-1/N$, which we adopt as the natural finite-size regularization. In this regime $v(c)>0$, the deterministic trajectory approaches $c=1$ monotonically, justifying the lower limit $c$. If one starts from a boundary state in a finite system, we set $c=1/N$; this choice only affects additive constants and not the leading scaling.

For $q = 1$, the drift $v(c)$ is linear, and the integral can be solved exactly, yielding [see Appendix~\ref{app:app_mot}]
\begin{equation}\label{eq:relax_linear}
    T(N, p) \sim \frac{1}{p} \ln N.
\end{equation}
This result shows that the consensus time grows logarithmically with the system size $N$ and is inversely proportional to the independence probability. In contrast, in the absence of an external field, the consensus time scales linearly with the system size $N$~\cite{redner2019reality}.

For $q > 1$, the integral yields the leading-order behavior
\begin{equation}\label{eq:relax_nonlinear}
    T(N, c(0), q, p) \approx \ln N + \mathcal{C}(c(0), q, p),
\end{equation}
where $\mathcal{C}(c(0), q, p)$ is a subleading correction term that is independent of $N$ but depends on the initial fraction $c(0)$, the nonlinear strength $q$, and the independence probability $p$. This correction captures the contribution from the global shape of the drift function $v(c)$ away from the absorbing boundary. In all cases, $\mathcal{C}(c(0), q, p) < 0$ for the admissible parameter range considered here.

In general, $\mathcal{C}(c(0), q, p)$ can be computed via partial fraction decomposition of the integrand $1/v(c)$ and expressed as:
\begin{equation}
    \mathcal{C}(c(0), q, p) = - \sum_{i = 1}^{q} \frac{\mathcal{A}_i}{r_i^2} \ln \left[1 - r_i \left(1 - c(0) \right) \right],
\end{equation}
where $r_i$ are the roots of $v(c)$ and the coefficients $\mathcal{A}_i$ arise from the decomposition. These constants are determined analytically for small values of $q$ and numerically for larger $q$. As an explicit example, for $q = 2$, the roots of the drift function $v(c)|_{q = 2}$ are $r_{1,2} = [3(1 - p) \pm \sqrt{(1 - p)(1 - 9p)}]/2$. In particular, for the balanced initial fraction $c(0) = 1/2$, the constant correction term $\mathcal{C}(p)$ simplifies to
\begin{align}\label{eq:relax_q23}
    \mathcal{C}(p) = &- \frac{3}{2} \sqrt{\dfrac{1-p}{1-9p}} \ln\left(\frac{1+3p-\sqrt{\left(1-p\right) \left(1-9p\right)}}{1+3p+\sqrt{\left(1-p\right) \left(1-9p\right)}}\right) \nonumber \\
    & - \frac{1}{2}\ln p.
\end{align}
Equation~\eqref{eq:relax_q23} also holds for $q = 3$, since both $q = 2$ and $q = 3$ have identical roots $r_i$.

\begin{figure*}[htbp]
    \centering
    \includegraphics[width=0.85\linewidth]{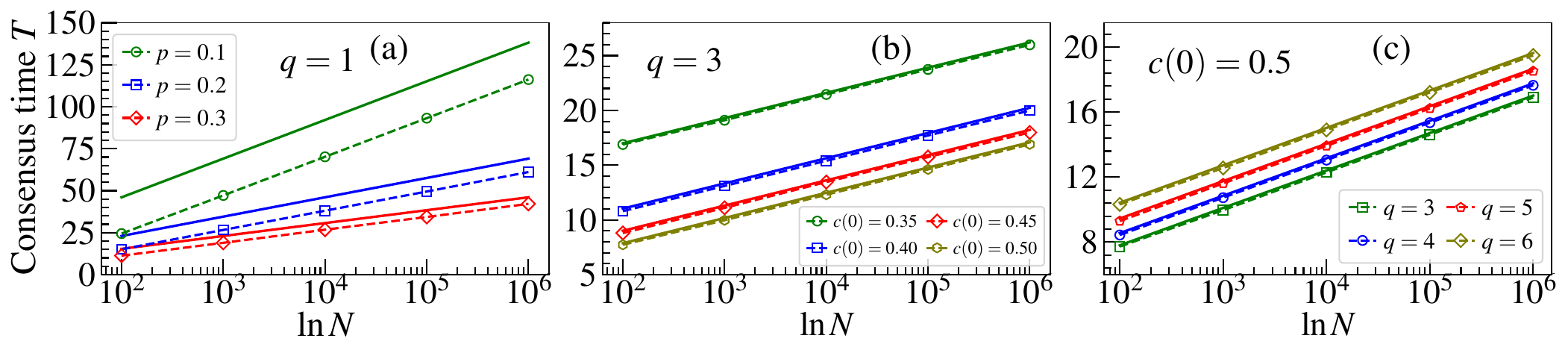}
    \caption{ Consensus time $T$ as a function of system size $N$ for various model configurations, with $s = 1$. (a) $q = 1$ with different values of $p$ at fixed initial fraction $c(0) = 0.5$. (b) $q = 3$ with various initial fraction $c(0)$ at $p = 0.1$. (c) Multiple values of $q$ at fixed $c(0) = 0.5$ and $p = 0.1$ for all $q > 1$. Solid lines represent the analytical predictions from Eq.~\eqref{eq:relax_linear} for $q = 1$, and Eq.~\eqref{eq:relax_nonlinear} for $ q > 1$, while dashed lines with markers correspond to MC simulation results, averaged over more than $10^4$ realizations.}
    \label{fig:relax_q3}
\end{figure*}

Figure~\ref{fig:relax_q3} presents a comparison between the theoretical predictions from Eqs.~\eqref{eq:relax_linear} and \eqref{eq:relax_nonlinear}, and results obtained from MC simulations.
It can be observed that the consensus time exhibits a leading-order logarithmic scaling concerning system size $N$ for all values of $q$, though the prefactor varies depending on the model parameters. Specifically, for $q = 1$, the prefactor scales as $1/p$, indicating that the system reaches consensus more rapidly as the independence probability increases. 

For all $q > 1$, the leading prefactor becomes independent of $q$ and takes a universal value of $1$. This prefactor coincides with that of the standard nonlinear $q$-VM under unbalanced initial conditions~\cite{ramirez2024ordering,kim2024competition}. Moreover, it differs from the prefactor of the standard nonlinear $q$-VM (corresponding to $p = 0$ in our framework) for balanced initial conditions, where it depends explicitly on the nonlinearity parameter $q$~\cite{kim2024competition}.

In general, the consensus time in the present model is shorter, i.e., has a smaller prefactor, than the standard $q$-VM for the same initial fraction $c(0)$. However, it is important to note that the consensus time of the standard $q$-VM cannot be directly recovered by simply setting $p = 0$ in our formulation. This is because the fixed-point structure of the dynamics in the present  model, including the positions of stable and unstable states, is inherently shaped by the presence of the independence term $p$. In contrast, the standard model assumes dynamics governed solely by group influence (nonlinear strength $q)$.

\section{Disordering Time}
The disordering time, refers to the characteristic timescale required for the system to transition from a consensus state to a disordered state. In the present model, disordering dynamics emerge when $s=\tfrac{1}{2}$ and $p > p_c(q)$ (with $q > 1$), indicating that under such conditions the system evolves toward a stable disordered state characterized by the vanishing of the order parameter, $m = 0$.

We employ Eq.~\eqref{eq:exit_approx} to analyze the disordering time by integrating from $c=1$ down to $c=1/2+1/\sqrt{N}$. Unlike the absorbing boundary at $c=1$, the symmetric point $c=1/2$ (for $s=1/2$) is a linearly stable interior fixed point when $p>p_c$ ($v'(1/2)<0$); hence the finite-size cutoff is set by the $\mathcal{O}(N^{-1/2})$ width of demographic fluctuations rather than by the lattice spacing. Accordingly, this cutoff incorporates the stochastic fluctuations around the stable fixed point and captures the scaling of the disordering time with the system size $N$. The choice is consistent with the natural fluctuation scale near $c=1/2$, $\delta c \sim \sqrt{\langle (c-1/2)^2\rangle}\sim N^{-1/2}$~\cite{van1992stochastic}. A similar finite-size regularization is commonly used in consensus-time analyses to avoid spurious trapping near unstable fixed points~\cite{kim2024competition}.

For a balanced system, where fluctuations dominate the dynamics, the average time required to transition from consensus to a disordered state scales as [see Appendix~\ref{appendix_F}]:
\begin{equation}\label{eq:dis_voter}
T(N, q, p) \sim \mathcal{B}(q, p) \ln N,
\end{equation}
with the prefactor $\mathcal{B}(q, p)$ given by (see Appendix~\ref{appendix_F}):
\begin{equation}\label{eq:prefactor_pol}
\mathcal{B}(q, p) = \left[ 2p - (1 - p)(q - 1) 2^{2 - q} \right]^{-1}.
\end{equation}
Notably, Eq.~\eqref{eq:dis_voter} indicates that the prefactor of the disordering time depends explicitly on both the nonlinear strength $q$ and the independence probability $p$, in contrast to the consensus time for $q > 1$, whose prefactor remains constant as shown in Eq.~\eqref{eq:relax_nonlinear}. It is worth emphasizing that Eq.~\eqref{eq:dis_voter} remains valid for all $p > p_c$ and $q > 1$.

Furthermore, the disordering time for the linear VM can be obtained directly from Eq.~\eqref{eq:dis_voter}, yielding:
\begin{equation}\label{eq:linear_polar}
T(N, p) \sim \frac{1}{2p} \ln N,
\end{equation}
which holds for all $p > 0$. Equation~\eqref{eq:linear_polar} reveals that the prefactor of the disordering time is half that of the consensus time in the linear case, as given in Eq.~\eqref{eq:relax_linear}. This result implies that, for a given $p$ and $N$, the system takes approximately half the time to reach a disordered state from an initial consensus compared to the time required to reach consensus from an initial disordered configuration.

\begin{figure*}[ht]
\centering
\includegraphics[width=0.85\linewidth]{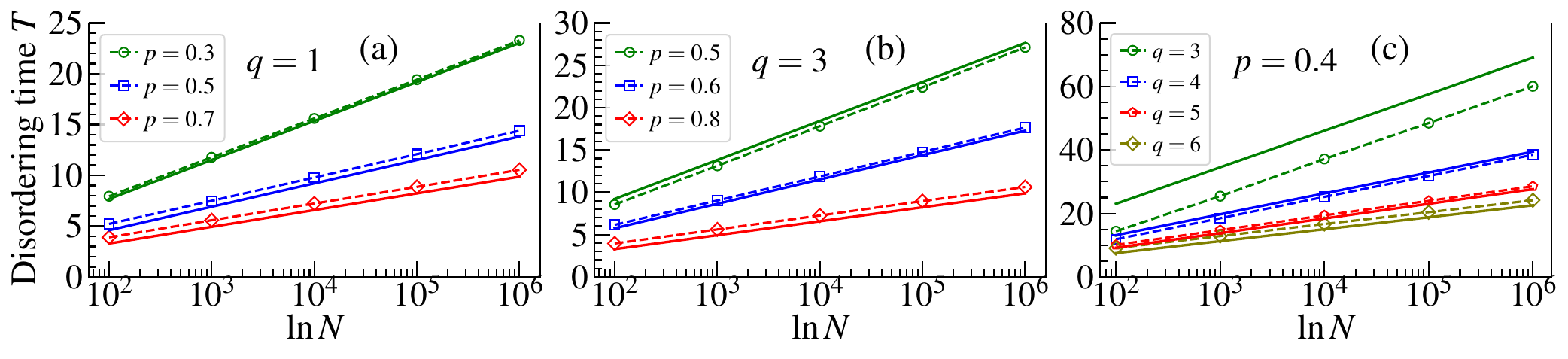}
\caption{Disordering time $T$ as a function of system size $N$ for various configurations of the model. (a) For $q = 1$ with several values of $p$. (b) For $q = 3$ with various values of $p$. (c) For multiple values of $q$ at fixed $p = 0.4$. Solid lines represent the analytical predictions from Eq.~\eqref{eq:dis_voter}, while dashed lines with markers correspond to MC simulation results, averaged over more than $10^4$ realizations.}
\label{fig:polar_q}
\end{figure*}

Figure~\ref{fig:polar_q} compares the analytical prediction of Eq.~\eqref{eq:dis_voter} with MC simulation results. The disordering time exhibits a leading-order logarithmic scaling with system size $N$, with a prefactor that depends on $p$ and $q$. As shown in panel (a) for $q = 1$ and panel (b) for $q = 3$, the disordering time decreases as $p$ increases, indicating faster disordering dynamics under stronger external influence. Panel (c), which summarizes results for various values of $q > 1$, demonstrates that for fixed $p$, increasing $q$ also reduces the disordering time, as consistently supported by both theoretical predictions and MC simulations.

The disordering time $T$ described by Eq.~\eqref{eq:dis_voter} diverges as $p$ approaches the critical probability $p_c$. In the regime $p > p_c$, the prefactor behaves as $|\epsilon|^{-\gamma}$, leading to the scaling form
\begin{equation}\label{eq:exit_time_pl}
T \sim |\epsilon|^{-\gamma} \ln N,
\end{equation}
with $\gamma = 1$, where $\epsilon = p$ for $q = 1$, and $\epsilon = p - p_c$ for $q > 1$.

This divergence is a hallmark of critical slowing down typically observed near absorbing-state phase transitions in mean-field systems~\cite{vazquez2008generic}. Near the critical point, the drift toward the disordered state diminishes substantially, causing stochastic fluctuations to take longer to overcome the stability barrier. This behavior is consistent with saddle–node bifurcations and critical dynamics, where the relaxation time diverges hyperbolically with a dynamic critical exponent of 1~\cite{hohenberg1977theory}. As $p$ increases further beyond $p_c$, the disordering process accelerates, and $T$ decreases, ultimately reaching its minimum value of $T_{\text{min}} = (\ln N)/2$ in the limit $ p\to 1$.

\begin{figure}[tb]
\centering
\includegraphics[width=0.6\linewidth]{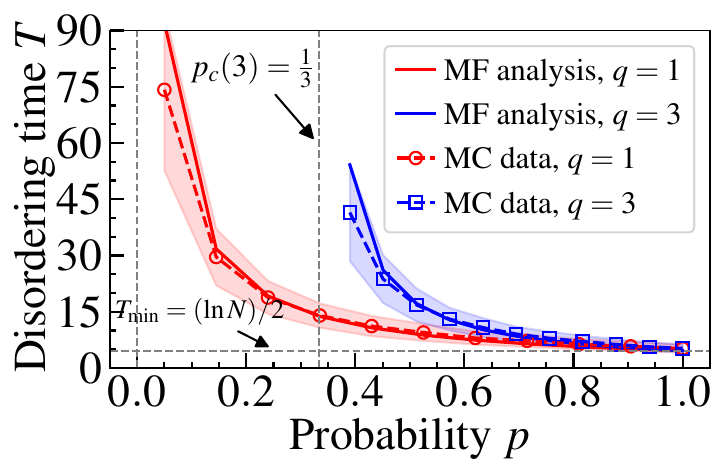}
\caption{Disordering time $T$ as a function of the independence probability $p$ for $q = 1$ and $q = 3$. Near the critical point $p_c$, the disordering time diverges and then decreases monotonically as $p$ increases, reaching the minimum value $T_{\text{min}} = (\ln N)/2$ in the limit $p \to 1$. Solid lines correspond to the analytical prediction from Eq.~\eqref{eq:dis_voter}, marker–dashed lines indicate MC simulation results, and shaded regions represent standard deviations. The system size is $N = 10^4$, and each data point is averaged over $10^4$ realizations.}
\label{fig:polar_vs_q}
\end{figure}

Using Eq.~\eqref{eq:dis_voter}, we analyze how the disordering time depends on the nonlinear strength $q$ under two distinct conditions. In the first scenario, the independence probability $p$ is kept fixed. In contrast, in the second, $p$ is tuned as a function of $q$ according to $p(q) = \alpha p_c(q)$, where $\alpha > 1$ is a constant that controls the distance between $p$ and the critical threshold $p_c(q)$ uniformly across all values of $q$, thereby ensuring that the system remains in the disordered regime, i.e., $p >p_c$ for all $q >1$.

In the first case, the disordering time exhibits a local maximum at a specific value $q^*$, given by
\begin{equation}\label{eq:optimum_point}
q^{*} = 1 + \frac{1}{\ln 2} \approx 2.44,
\end{equation}
which marks a bottleneck in the dynamics, i.e., the slowest disordering occurs when the nonlinear strength $q$ is approximately 2–3. As $q$ increases toward $q^*$, local conformity becomes more dominant, delaying the formation of competing opinion clusters. For $q > q^*$, the influence of independence grows stronger, accelerating the disordering process. Eventually, as $q$ increases, the disordering time saturates and approaches an asymptotically minimal value, typically for $q \gtrsim 8$.

In the second case, when $p$ is tuned proportionally to the critical threshold, $p = \alpha p_c(q)$, substituting the tuned parameter into Eq.~\eqref{eq:dis_voter} yields
\begin{equation}\label{eq:T_tunned}
T(N, q, \alpha) \sim \dfrac{2^{q-2}}{\left(\alpha - 1\right)\left(q-1\right)} \ln N.
\end{equation}
Under this parametrization, the local extremum at $q^*$ becomes a local minimum, indicating a ``sweet spot" for the fastest disordering dynamics. At this point, the trade-off between local conformity and externally tuned independence achieves optimal conditions for fragmentation of opinions near the critical regime. Interestingly, the value $q^*$ is independent of the value of $p$. suggesting that this optimality is an intrinsic feature of the standard nonlinear $q$-VM.

\begin{figure}[tb]
\centering
\includegraphics[width=\linewidth]{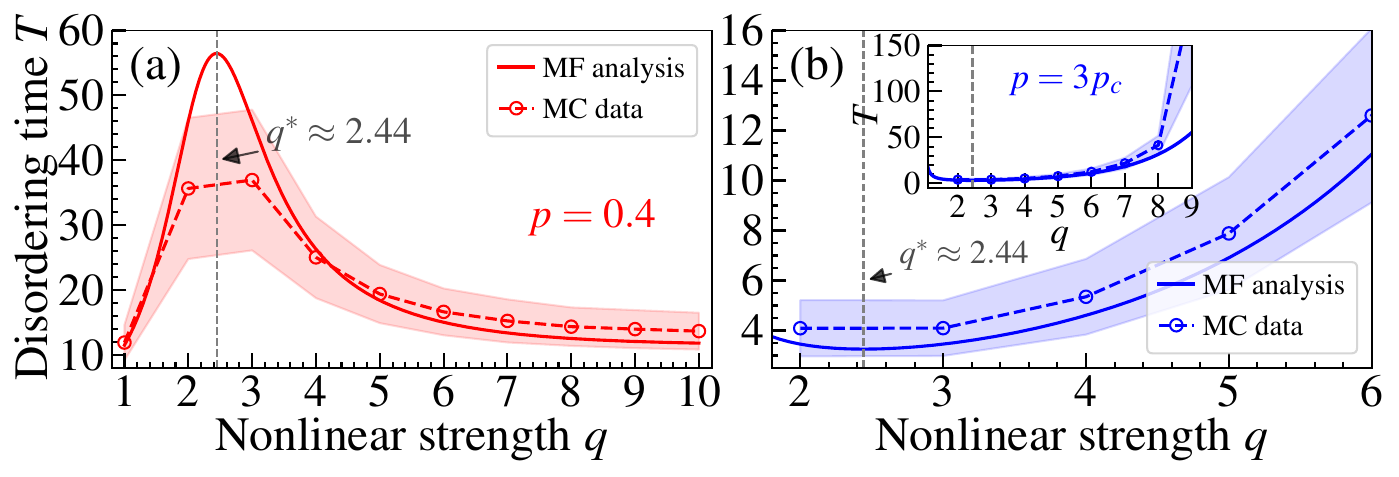}
\caption{Disordering time $T$ as a function of the nonlinearity parameter $q$ under two conditions. (a) Fixed independence probability $p = 0.4$. (b) Tuned independence probability $p = \alpha \,p_c(q)$ with $\alpha = 3$. A nonmonotonic behavior is observed in both cases: a local maximum in panel (a) and a local minimum in panel (b), both located at the optimal point $q^* \approx 2.44$. Solid lines represent analytical predictions from Eq.~\eqref{eq:dis_voter} and Eq.~\eqref{eq:T_tunned}, respectively. Marker–dashed lines denote MC simulation results, and shaded areas indicate standard deviations. System size is $N = 10^4$ for panel (a) and $10^3$ for panel (b), with each data point averaged over $10^4$ realizations.}
\label{fig:polar_q_optimum}
\end{figure}

\section{Exit probability}\label{sec:VI}
We define the exit probability to the all-$+1$ consensus as $E(c(0)) \equiv \Pr\{\tau_1 < \tau_0 \mid c(0)\}$, where $\tau_0$ and $\tau_1$ are the first-passage times to $c=0$ and $c=1$, respectively, under absorbing boundaries at $c = 0$ and $c = 1$. By symmetry, the probability to exit at $c=0$ is $1-E(c(0))$. This standard definition requires absorbing states. Accordingly, throughout this section we consider $p=0$ (the field term vanishes and the results are independent of $s$), and $p>0$ with extreme biases $s\in\{0,1\}$, for which one of the consensus states is absorbing. For $p>0$ and intermediate bias $0<s<1$, neither boundary is absorbing and the usual exit probability is not defined. In this regime one may instead study first-passage statistics with artificial absorbing boundaries or quasi-stationary switching between metastable states, which lie outside our scope here. Unless stated otherwise, simulations start from the initial fraction $c(0)$, and parameter values $(q,p,s)$ are indicated in the figure legends.

The analytical treatment of the exit probability is based on a second-order Kramers--Moyal expansion of the Fokker--Planck equation, as discussed in Appendix~\ref{app:app_exip}. The resulting backward Kolmogorov equation admits an exact integral solution given by:
\begin{equation}\label{eq:exit_pp}
    E(c(0)) = \dfrac{\displaystyle \int_{0}^{c(0)} \exp\left[ -\int_{0}^{y} \frac{v(u)}{D(u)}\, du\right] dy}{\displaystyle \int_{0}^{1} \exp\left[ -\int_{0}^{y} \frac{v(u)}{D(u)}\, du\right] dy}.
\end{equation}

In the case of the linear VM, this integral can be evaluated explicitly to yield
\begin{widetext}
\begin{equation}\label{eq:exit_p_ori}
    E(c(0)) = 
\begin{cases} 
\dfrac{\left[ p + 2c(0)\left(1 - p\right) \right]^{\eta} - p^{\eta}}{\left(2 - p\right)^{\eta} - p^{\eta}}, & \text{for} \quad s = 1, \\[8pt]
\dfrac{\left[ \left(2 - p\right) - 2c(0)\left(1 - p\right) \right]^{\eta} - \left(2 - p\right)^{\eta}}{p^{\eta} - \left(2 - p\right)^{\eta}}, & \text{for} \quad s = 0,
\end{cases}
\end{equation}
\end{widetext}
where $\eta = 1 - Np/(1 - p)$. For $p = 0$, the exit probability reduces to the well-known result for the classical linear VM, namely $E(c(0)) = c(0)$, which is independent of the system size $N$~\cite{redner2019reality}. 

% The exit probability $E(c(0))$ in Eq.~\eqref{eq:exit_p_ori} characterizes the likelihood that, starting form initial fraction $c(0)$, the process hits $c =1$ before $c = 0$ under absorbing boundaries at both ends (i.e., a splitting probability), incorporating both finite-$N$ fluctuations and the drift induced by the field term. For $s = 1$, the term $\left[p + 2c(0)\left(1-p\right)\right]^\eta$ encapsulates the effective contribution from local majority-rule interactions, nonlinearly amplified by the exponent $\eta$, while the subtraction of $p^\eta$ isolates the purely stochastic component. The normalization factor $\left[(2 - p)^\eta - p^\eta\right]$ ensures that $E(c(0))$ remains confined within the unit interval, reflecting a competition between disorder (noise) and order (interactions) that is typical in systems near a critical threshold.

The exit probability $E(c(0))$ in Eq.~\eqref{eq:exit_p_ori} characterizes the likelihood that, starting form the initial fraction $c(0)$, the process hits $c =1$ before $c = 0$ under absorbing boundaries at both ends (i.e., a splitting probability), incorporating both finite-$N$ fluctuations and the drift induced by the field term. For $s = 1$, the term $\left[p + 2c(0)\left(1-p\right)\right]^\eta$ arises from the backward Kolmogorov solution with $v(c) = p \left(1-c\right)$ and $D(c) \propto c\left(1-c\right)/N$. The subtraction of $p^{\eta}$ enforces the boundary condition at $c = 0$. The normalization factor $\left[(2 - p)^\eta - p^\eta\right]$ ensures that $E(c(0))$ remains within $[0,1]$ and fixes $E(0) = 0$, and $E(1) = 1$, encoding the balance between drift toward the favored state and finite-size diffusion. Conversely, for $s = 0$, the expression follows from bias–inversion symmetry, $E_{s=0}(c(0)) = 1 - E_{s=1}(1 - c(0))$; equivalently, it can be obtained from the $s = 1$ formula by the substitutions $c(0) \to 1-c(0)$ and $p \to 2-p$. The exponent $\eta$ arises from the backward equation through the ratio $v/D$ and encodes the finite-$N$ drift-diffusion balance; it does not signal critical behavior.

% Conversely, for $s = 0$, the functional form is modified by replacing $p$ with $ 2-p$, effectively inverting the direction of the external bias. This symmetry between the two cases underscores the system's sensitivity to microscopic parameters and the reversibility of its macroscopic behavior under bias inversion. The nonlinear structure introduced by the exponent $\eta$ is reminiscent of critical phenomena, in which small changes in local dynamics can drive large-scale transitions between distinct phases. 

\begin{figure}[tb]
    \centering
    \includegraphics[width=\linewidth]{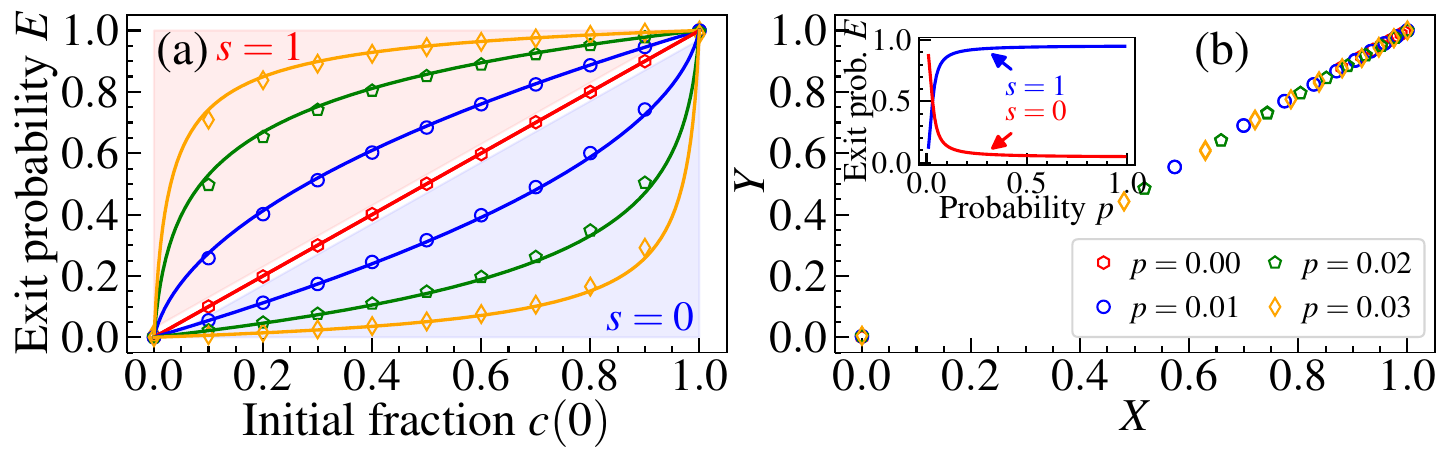}
    \caption{Exit probability of the linear VM for various values of the independence probability $p$, with system size fixed at $N = 50$. Panel (a) shows the standard plot of $E(c)$, while panel (b) presents a scaling collapse using the variables defined in Eqs.~\eqref{eq:x_scaling} and~\eqref{eq:y_scaling}. The inset in panel (b) displays the exit probability as a function of $p$ for $c = 0.03$ with $s = 1$ (red-shaded region), and $c = 0.97$ with $s = 0$ (blue-shaded region). Solid lines correspond to the analytical prediction in Eq.~\eqref{eq:exit_p_ori}, while symbols represent MC simulation results averaged over more than $10^4$ realizations.}
    \label{fig:exit_voter}
\end{figure}

Figure~\ref{fig:exit_voter}(a) illustrates that increasing the independence probability $p$ enhances the exit probability. The Monte Carlo results agree well with the analytical prediction of Eq.~\eqref{eq:exit_p_ori}, even for relatively small system size $N$. As seen in Fig.~\ref{fig:exit_voter}(b), the exit probability can be recast into a universal scaling form independent of both the independence probability $p$ and the system size $N$. Specifically, for the $ s=1$ case, the exit probability is given by
\begin{equation}
    E(c(0)) = \frac{\left[p + 2\left(1 - p\right)c(0)\right]^\eta - p^\eta}{(2 - p)^\eta - p^\eta}.
\end{equation}
By introducing the scaled variable
\begin{equation}\label{eq:x_scaling}
    X = \frac{\ln\left[\left(p + 2\left(1 - p\right)c(0)\right)/p\right]}{\ln\left[(2 - p)/p\right]},
\end{equation}
and the corresponding scaled exit probability
\begin{equation}\label{eq:y_scaling}
    Y = \frac{\ln\left\{1 + \left[\left(\frac{2 - p}{p}\right)^\eta - 1\right]E(c(0))\right\}}{\eta \ln\left[\left(2 - p\right)/p\right]},
\end{equation}
one obtains the linear relation $Y = X$, which defines a universal master curve. Importantly, since the $s = 0$ case is related via the symmetry $E_{s=0}(c(0)) = 1 - E_{s=1}(1 - c(0))$, the same scaling variables apply upon substituting $c(0) \to 1 - c(0)$ and $E \to 1-E$. As a result, for arbitrary values of $p$ and $N$, the entire family of exit probability curves collapses onto a single straight line $Y = X$ in the $(X, Y)$ plane, revealing the underlying universality of the system's dynamics.

In the absence of independence $(p=0)$ and for nonlinear strength $q > 1$, the exit probability in Eq.~\eqref{eq:exit_pp} can be approximated using the saddle-point (Laplace) method as:
\begin{equation}\label{eq:exit_general_eff}
    E(c(0),q,N) \approx \dfrac{1}{2} \left[1+ \mathrm{erf}\left(\sqrt{2N_{\text{eff}}(q-1)} \left(c(0) - \tfrac{1}{2} \right)\right)\right],
\end{equation}
where $N_{\text{eff}} = N/q$. While the original derivation suggests a dependence on the total system size $N$, numerical simulations reveal that the correct scaling is obtained by replacing $N$ with an effective population size $N_{\text{eff}} = N/q$. This adjustment reflects that, although there are $N$ agents in the system, opinion updates are governed by local interactions within groups of size $q$. As such, the number of statistically independent units driving the macroscopic evolution scales as $N/q$, effectively reducing the system's degrees of freedom. Similar scaling behaviors have been observed in related models involving group-based update rules~\cite{castellano2009nonlinear, jkedrzejewski2017pair}.

Moreover, although the nonlinearity parameter $q$ enters the scaling factor as $\sqrt{(q - 1)/q}$, this dependence becomes negligible as $q$ increases, with the prefactor quickly approaching unity. Consequently, in the large-$q$ limit, the exit probability becomes effectively independent of $q$, with its macroscopic profile dominated by the system size and the initial fraction $c(0)$. This saturation phenomenon implies that increasing $q$ beyond a certain threshold no longer significantly alters the collective behavior. For example, defining saturation as a relative deviation in $E(c(0),q,N)$ below 5\%, this regime is already reached for $q \geq 10$.

\begin{figure}[tb]
    \centering
    \includegraphics[width=\linewidth]{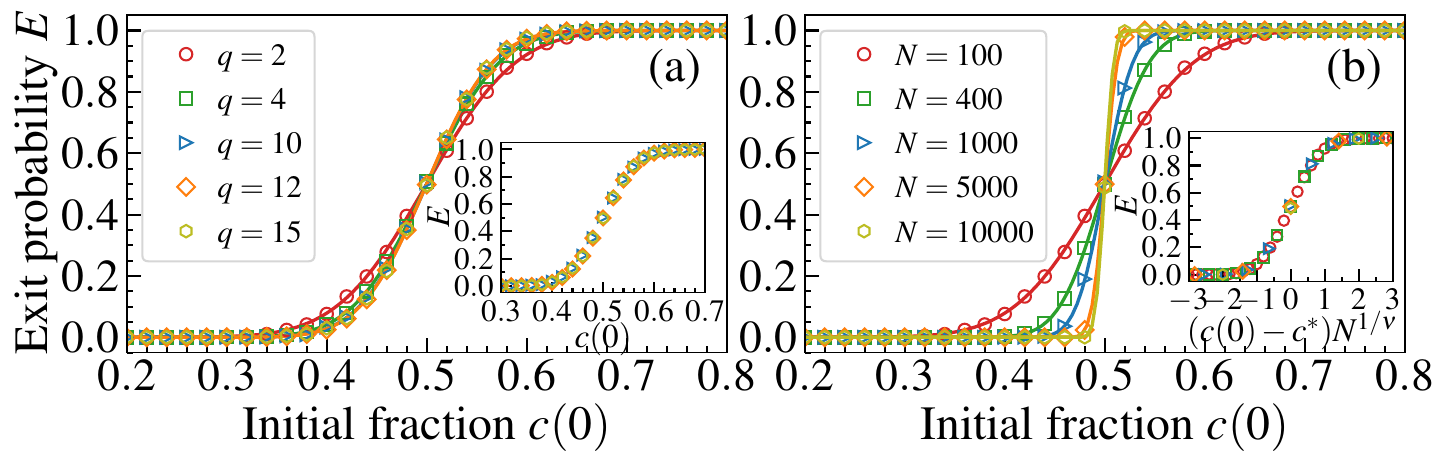}
    \caption{Exit probability of the $q$-voter model at $p=0$ (independent of $s$). Panel (a) shows results for fixed $N=100$ and various $q$. The inset highlights the saturation behavior for large $q$ ($q\gtrsim 10$), where the curves are nearly indistinguishable. Panel (b) presents results for fixed $q=2$ and varying system size $N$. The inset illustrates finite-size scaling with symmetry point $c^*=1/2$ and effective exponent $\nu\simeq 2$. In both panels, solid lines denote the analytical prediction from Eq.~\eqref{eq:exit_general_eff}, while symbols are MC results averaged over more than $10^4$ realizations.}
    \label{fig:exit_prob}
\end{figure}

The prediction of Eq.~\eqref{eq:exit_general_eff} for various values of $q$ shows excellent agreement with MC simulation results, as illustrated in Fig.~\ref{fig:exit_prob}(a). The inset in panel (a) displays the data for $q = 10$, $12$, and $15$, which visually overlap and become increasingly indistinguishable, thereby providing clear evidence of the saturation behavior for $q \geq 10$ within a $5\%$ tolerance.

Panel (b) of Fig.~\ref{fig:exit_prob} presents the exit probability $E(c(0))$ for a fixed nonlinearity $q = 2$ and varying system sizes $N$, highlighting the dependence of the transition sharpness on $N$. As the system size increases, the transition from $E(c(0)) \approx 0$ to $E(c(0)) \approx 1$ becomes progressively steeper and increasingly localized around the critical point $c^* = 1/2$, on a characteristic scale $\Delta c \sim 1/\sqrt{N}$. The inset in panel (b) demonstrates the corresponding finite-size scaling behavior: when the data are plotted against the rescaled variable $x = \sqrt{2N\left(q-1\right)/q}\,\left(c(0) - \frac{1}{2}\right)$, they collapse neatly onto the universal curve $E(x) \approx \frac{1}{2}\left[1+\operatorname{erf}(x)\right]$, validating the scaling form.

This observed scaling behavior is fully consistent with a finite-size scaling description, which states that, near the symmetry point $c^*$, relevant observables depend solely on the scaling combination $(c(0) - c^*)N^{1/\nu}$~\cite{cardy1996scaling, stanley1971phase}. Here $\nu = 2$ should be understood as an effective finite-size exponent governing the width of the crossover, which in our case is $\nu \simeq 2$. The reference point is $c^* = 1/2$ (the symmetry, which is unstable in the nonlinear case). As a result, numerical data obtained for various system sizes $N$ and different values of $q$, when plotted against the scaling variable $x$, collapse onto a single universal curve. This data collapse supports the validity of the scaling description and confirms the asymptotic analytical picture of the system’s behavior in the vicinity of the saddle (unstable) point $c^*$.

For $p>0$, a straightforward quadratic saddle-point (Laplace) expansion can become inadequate whenever the dominant contribution is not a nondegenerate interior saddle. For example, when the stationary point $c^*$ approaches a boundary (endpoint dominance), or near fold (saddle–node) loci where two stationary points coalesce and $|v' (c^*)|\to 0$. In these regimes, the contribution to the integral is no longer confined to an infinitesimally narrow, symmetric neighborhood of a single saddle, and higher-order terms beyond quadratic become significant. In particular, when $s\neq\tfrac12$ the condition $v(c;p,s,q)=0$ is satisfied at a shifted $c^*\neq 1/2$ (whereas for $s=\tfrac12$ one has $c^*=1/2$ away from bifurcation lines). As $|v' (c^*)|$ diminishes near a fold, the effective contributing region broadens and the quadratic expansion loses accuracy. In such cases, one should use uniform asymptotics (e.g., boundary-layer, or Airy-type approximations for coalescing saddles) to capture the global behavior beyond the local vicinity of $c^*$.

% For sufficiently large values of $p$, the standard saddle-point (Laplace) approximation becomes inadequate, as the dominant contributions to the integral are no longer confined to an infinitesimally narrow and symmetric neighborhood around a single saddle point. In particular, for $p > 0$, the condition $v(c; p) = 0$ is satisfied at a shifted position $c^*$ that deviates from the symmetric point $ c^* = 1/2$, valid for $p = 0$. Moreover, the reduced local curvature $v'(c^*)$ broadens the region contributing significantly to the integral. Consequently, the conventional quadratic expansion around the saddle point $c^*$ becomes inaccurate, neglecting essential higher-order terms that considerably affect the asymptotic evaluation. Intuitively, the transition region broadens and becomes more diffuse, violating the fundamental assumption of local dominance required by the saddle-point method. Consequently, more sophisticated uniform asymptotic techniques are required to accurately capture the system's global behavior beyond the local vicinity of $c^*$.

\begin{figure}[tb]
    \centering
    \includegraphics[width=\linewidth]{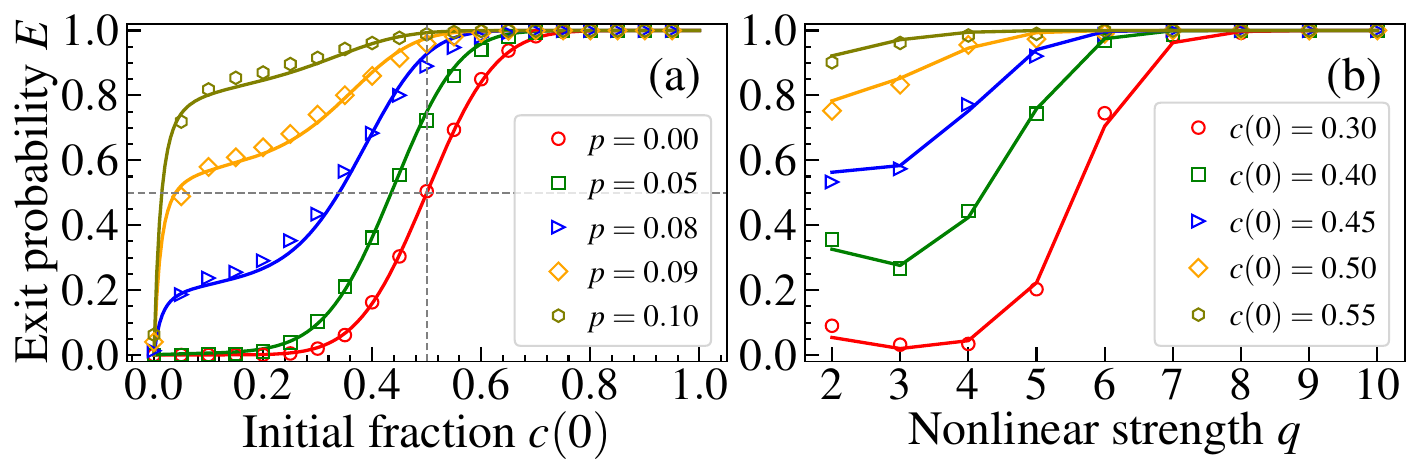}
    \caption{Exit probability of the nonlinear VM for $p > 0$ and $s = 1$. Panel (a) shows results for fixed $q = 2$ and several values of the independence probability $p$. Panel (b) shows results for fixed $p = 0.05$ and various initial fraction $c(0)$. In both panels, the system size is fixed at $N = 50$. Solid lines represent numerical solutions of Eq.~\eqref{eq:exit_pp}, while symbols denote MC simulation results averaged over more than $10^4$ realizations.}
    \label{fig:exit_prob_p}
\end{figure}

Figure~\ref{fig:exit_prob_p} compares the numerical solutions of Eq.~\eqref{eq:exit_pp} with MC simulations, showing good quantitative agreement across various parameters. Panel (a) demonstrates that the tendency of the exit probability $E$ to approach the fully ordered $+1$ state becomes increasingly pronounced as $p$ increases. This observation aligns with the physical interpretation of the probability $p$ when the bias parameter is fixed at $s = 1$: larger values of $p$ enhance the influence of the external field, thereby driving the system toward the consensus of the $+1$ state. Panel (b) displays the exit probability $E$ as a function of the nonlinear strength $q$ for various initial conditions $c(0)$. For small $q$, particularly when $c < 0.5$, the system exhibits a pronounced nonmonotonic behavior due to the competition between the external field, which promotes consensus toward the $+1$ state, and nonlinear local interactions favoring the $ 1$ state. This competition is most visible for $c(0) = 0.30$ and $c(0) = 0.40$, where $E$ first increases and then sharply rises as $q$ becomes large enough to suppress the effect of local fluctuations. As $q$ increases, the influence of the external field becomes dominant, and $E$ monotonically approaches 1, indicating a transition toward a consensus state. This saturation occurs for $q \gtrsim 8$, beyond which further increases in $q$ do not significantly alter the outcome. In contrast, for $c \geq 0.50$, the system already favors the $+1$ consensus, and $E$ remains close to 1 regardless of $q$, confirming the strong biasing effect of the external field under favorable initial conditions.

\section{Summary and Conclusion}
We investigated a variant of the voter model incorporating local interactions and a random external field. In this model, agents adopt the unanimous opinion of a randomly selected group of $q$ agents with probability $1-p$, or, with probability $p$, independently align with an external field by taking state $+1$ with probability $s$ (and $-1$ otherwise). Using a mean-field approximation, we identify an order–disorder phase transition at a critical independence probability $p_c$ that depends on the nonlinear strength $q$, occurring only at the symmetric bias $s=\tfrac{1}{2}$. Deviations from symmetry break the up–down invariance and generally drive the system toward the favored state; for small asymmetry $\delta s\equiv |s-\tfrac{1}{2}|\ll1$, multistability can still occur.

Analyzing the evolution of the fraction $c$ of agents in state $+1$, we found distinct behaviors depending on $q$: for $q=1$, the stationary fraction equals the external bias ($ c \to s$). For $q > 1$, spontaneous ordering emerges contingent on $p$ and $s$. For $s\neq\tfrac12$, the bias breaks the up–down symmetry and drives the system toward the favored state; however, for small asymmetry $\delta s\ll1$ and suitable $p$, multiple stable steady states can coexist (cusp for $1<q<5$ and butterfly for $q>5$), whereas sufficiently large $\delta s$ leaves a single stable state. At the symmetric point $s=\tfrac12$, the steady-state outcome is highly sensitive to $p$: for $p<p_c$, the system approaches $c=\tfrac12\pm\delta$ with $\delta$ depending on $p$ and $q$, whereas for $p>p_c$ it evolves to the symmetric state $c=\tfrac12$. (In particular, the transition is continuous for $1<q<5$ and discontinuous for $q>5$, corresponding to supercritical and subcritical pitchforks, respectively.)

We further analyzed the consensus time $T$ required to reach ordered states. For extreme biases, we demonstrated that consensus time scales logarithmically with system size $N$ as $T \sim \mathcal{B}\ln N$, where $\mathcal{B} = 1/p$ for $q = 1$ and $\mathcal{B} = 1$ for $q > 1$, highlighting fundamental differences between linear and nonlinear strength. These analytical predictions align closely with MC simulations. At the symmetric bias, disordering time also scales logarithmically as $T \sim \mathcal{B}\ln N$, where $\mathcal{B} = 1/(2p)$ for $q = 1$, and for $q > 1$, $\mathcal{B}$ depends explicitly on $p > p_c$ and $q$. We identified a specific nonlinear strength $q^* \approx 2.44$ ($q$ between $2$ and $3$), independent of $p$, which maximizes the disordering time. Additionally, we found a universal minimum disordering time $T_{\text{min}} = \ln N/2$ occurring at maximal independence probability $p$.

Finally, we examined the exit probability $E(c(0))$, defined as the probability that, starting from an initial fraction $c(0)$ (e.g., $c = 1$), the process reaches a specified absorbing consensus state. For the linear case ($q = 1)$, $E(c(0))$ has a closed form that deforms smoothly with $p$ and reduces to $E(c(0)) = c(0)$ at $p = 0$. In contrast, for $q > 1$ at $p = 0$, $E(c(0))$ exhibits a sharp transition at $c = 1/2$, approaching a Heaviside step function as system size $N \to \infty$. For $p > 0$ and intermediate bias $0 < s < 1$, the standard splitting probability with two absorbing boundaries is not defined; for extreme bias $s \in \{0,1\}$ and $ p > 0$ absorption at the favored boundary occurs with probability one. For large $q$, the $p=0 $ curves saturate and become nearly indistinguishable.

In summary, our results elucidate how external bias and local interaction rules jointly control ordered and disordered phases in voter dynamics, providing analytical insight into consensus formation and disorder onset under random external bias.

% In summary, our results elucidate how external bias and local interaction rules collectively dictate the emergence of ordered and disordered phases in voter dynamics, providing analytical insight into consensus formation and disorder onset under random external bias.

%%%%%%%%%%%%%%%%%%%%%%%%%%%%%%%%%%%%%%%%%%%%%%%%%%%%%%%%%%%%%%%%%%%%%%%%%%%%%%%%
% \section*{Declaration of Interests}
% The contributors declare that they have no apparent competing business or personal connections that might have appeared to have influenced the reported work.

%%%%%%%%%%%%%%%%%%%%%%%%%%%%%%%%%%%%%%%%%%%%%%%%%%%%%%%%%%%%%%%%%%%%%%%%%%%%%%%%
\section*{Acknowledgments}
\textbf{R.~Muslim} was supported by the Asia Pacific Center for Theoretical Physics, funded by the Science and Technology Promotion Fund and Lottery Fund of the Korean Government, as well as by the Management Talent Program of the National Research and Innovation Agency of Indonesia (BRIN). \textbf{J. Kim} was supported by the National Research Foundation of Korea (NRF) under Grants No. 2020R1A2C2003669 and No. RS-2025-00558837.

% \onecolumngrid
% \newpage
\appendix
\section{\label{appendix_A} Transition Probability}

In this section, we present the derivation of the microscopic transition probabilities governing the evolution of the number of agents in the up-state ($+1$) during the discrete-time dynamics of the model. At each time step $\delta t = 1/N$, a single agent is randomly selected and may change its state, resulting in a possible change of the global order parameter $c$ by an amount $\delta c = 1/N$. Thus, the system can evolve by increasing, decreasing, or maintaining the current number of up-state agents.

The update rule is defined as follows: with probability $p$, the selected agent acts independently of its neighbors and responds to a random external field. In this case, a voter in state $-1$ switches to $+1$ with probability $s$, while a voter in state $+1$ switches to $-1$ with probability $ 1-s$. With probability $ 1-p$, the agent follows the opinion of a randomly selected group of $q$ neighbors and adopts their unanimous opinion if unanimity is present.

Accordingly, the probability that a voter in state $-1$ switches to $+1$ due to either external influence or unanimous conformity is given by the raising operator $R$. Conversely, the probability of a transition from $+1$ to $-1$ is described by the lowering operator $L$. These are defined as
\begin{align}
    R  &= (1 - p)\, \frac{N_{\downarrow} \prod_{i=1}^{q} (N_{\uparrow} - i + 1)}{\prod_{i=1}^{q+1} (N - i + 1)} + p s\, \frac{N_{\downarrow}}{N},  \label{eq:app_discrete_R}\\
    L  &= (1 - p)\, \frac{N_{\uparrow} \prod_{i=1}^{q} (N_{\downarrow} - i + 1)}{\prod_{i=1}^{q+1} (N - i + 1)} + p(1 - s)\, \frac{N_{\uparrow}}{N}, \label{eq:app_discrete_L}
\end{align}
where $N_{\uparrow}$ and $N_{\downarrow}$ denote the number of agents in the up and down states, respectively, such that $N = N_{\uparrow} + N_{\downarrow}$ is the total number of agents in the system.

In the thermodynamic limit $N \to \infty$, it is convenient to introduce the continuous variable $c = N_{\uparrow}/N$ representing the fraction of up-state agents. In this continuum approximation, Eqs.~\eqref{eq:app_discrete_R} and \eqref{eq:app_discrete_L} reduce to
\begin{align}
    R(c) &= (1 - c)\left[(1 - p)\, c^q + p s\right], \label{eq:app_continuous_R} \\
    L(c) &= c \left[(1 - p)\, (1 - c)^q + p(1 - s)\right], \label{eq:app_continuous_L}
\end{align}
which correspond to Eqs.~\eqref{eq:vara_12} and \eqref{eq:varb_13} in the main text. 
\section{\label{appendix_B} Time Evolution and Stationary Condition}

In the thermodynamic limit $N \to \infty$, the time evolution of the fraction $c(t)$ of agents in the up-state can be described deterministically via the net transition rate. The evolution equation is given by
\begin{equation}\label{eq:time_int_app}
    \int_{c(0)}^{c(t)} \frac{du}{R(u) - L(u)} = t,
\end{equation}
where $R(u)$ and $L(u)$ are the continuous raising and lowering transition probabilities defined in Eqs.~\eqref{eq:vara_12} and \eqref{eq:varb_13} of the main text. In many cases, the difference $R(u) - L(u)$ can be written in a factorized polynomial form,
\begin{equation}\label{eq:poly_K}
    R(u) - L(u) = -K \prod_{i=1}^{n} (u - r_i),
\end{equation}
where $K > 0$ is a constant depending on $p$ and $q$, and $\{r_i\}$ are the real or complex roots of $R(u) = L(u)$, corresponding to fixed points of the dynamics.

Let $f(u) = u(1 - u)[u^{q-1} - (1 - u)^{q - 1}]$ denote the nonlinear contribution to $R(u) - L(u)$. For $q = 1$, $f(u) = 0$, and the transition rate reduces to a linear function $R(u) - L(u) = -p\left(u-s\right)$ with $K = p$ and a single root $r_1 = s$. For even $q > 1$, the leading order of the nonlinear term is of degree $q + 1$, with coefficient $K = 2\left(1 - p\right)$. For odd $q > 1$, the highest-order terms cancel by symmetry, reducing the degree to $q$ with $K = (q - 1)(1 - p)$.

Substituting Eq.~\eqref{eq:poly_K} into Eq.~\eqref{eq:time_int_app}, and assuming all roots are distinct, yields the formal solution
\begin{align}
    \sum_{i=1}^{n} \frac{1}{\prod_{j \neq i} (r_i - r_j)} \ln \left| \frac{c(t) - r_i}{c(0) - r_i} \right| = -K t,
\end{align}
or equivalently in exponential form,
\begin{equation}
    \prod_{i=1}^{n} \left| \dfrac{c(t) - r_i}{c(0) - r_i} \right|^{\dfrac{1}{\prod_{j \neq i} (r_i - r_j)}} = \exp(-K t),
\end{equation}
which provides an implicit solution for the time evolution of $c(t)$, as presented in Eq.~\eqref{eq:implicit} of the main text.

For $q = 1$, the dynamics reduce to a simple exponential relaxation:
\begin{equation}
    \frac{c(t) - s}{c(0) - s} = \exp(-p t), \implies c(t) = s + \left[c(0) - s\right] \exp(-p t).
\end{equation}
For $q = 2, 3$, the implicit solution consists of a sum of logarithmic terms over three roots $\{r_1, r_2, r_3\}$ of the cubic equation:
\begin{align}
    & \sum_{i=1}^{3} \frac{1}{\prod_{j \neq i} (r_i - r_j)} \ln \left| \frac{c(t) - r_i}{c(0) - r_i} \right| = -2(1 - p)t.
\end{align}
The roots are obtained from the cubic equation
\begin{equation}\label{eq:cubic_poly}
    u^3 - \frac{3}{2} u^2 + \frac{1}{2(1 - p)} u - \frac{p s}{2(1 - p)} = 0.
\end{equation}
Defining the shifted variable $z = u - 1/2$, Eq.~\eqref{eq:cubic_poly} becomes the depressed cubic
\begin{equation}
    z^3 + \mathcal{P} z + \mathcal{Q} = 0,
\end{equation}
with
\begin{equation}
    \mathcal{P} = \frac{1 - 3p}{4(1 - p)}, \qquad \mathcal{Q} = \frac{p(1 - 2s)}{4(1 - p)}.
\end{equation}
The three roots $r_i = z_i + 1/2$ are then given explicitly by
\begin{equation}
    r_i = \frac{1}{2} + 2 \sqrt{\frac{\mathcal{P}}{3}} \cos \left[ \frac{1}{3} \arccos \left( \frac{-\mathcal{Q}}{2} \sqrt{ \frac{27}{\mathcal{P}^3} } \right) - \frac{2\pi(i - 1)}{3} \right].
\end{equation}

In the limit $p \to 0$, one obtains $\mathcal{P} = 1/4$ and $\mathcal{Q} = 0$, yielding the roots $r_1 = 1$, $r_2 = 1/2$, and $r_3 = 0$, which correspond to two stable fixed points and one unstable saddle, consistent with the behavior of the deterministic $q$-VM without bias.

The stationary state of the system satisfies the condition $dc/dt = 0$, or equivalently $R(c) = L(c)$. This condition leads to the equation
\begin{equation}\label{eq:fixed_point_cond}
    (1 - p) \left[ c^q - c(1 - c)^q - c^{1 + q} \right] - p\left(c - s\right) = 0.
\end{equation}
Solving Eq.~\eqref{eq:fixed_point_cond} for $c$ is generally intractable for arbitrary $q$. However, it is more convenient to express it as an explicit function for $p$:
\begin{equation} \label{eq:pc_explicit}
    p(c, q, s) = \frac{c(1 - c)^q + c^{1 + q} - c^q}{c^{1 + q} + c(1 - c)^q - c^q - c + s},
\end{equation}
which corresponds to Eq.~\eqref{eq:stat} in the main text.

Evaluating Eq.~\eqref{eq:pc_explicit} in the limit $c \to 1/2$ yields the critical point of the ordering–disordering transition. Since both the numerator and denominator vanish at this point, applying L’Hôpital’s rule gives
\begin{equation}\label{eq:pc_result}
    p_c = \lim_{c \to 1/2} \frac{f'(c)}{g'(c)} = \frac{q - 1}{q - 1 + 2^{q - 1}},
\end{equation}
where $f(c)$ and $g(c)$ denote the numerator and denominator of Eq.~\eqref{eq:pc_explicit}, respectively. This expression defines the critical threshold for the independence probability that separates the ordered and disordered phases when the bias $s$ is symmetric.

\section{\label{appendix_C}Stability and Instability of Fixed Points}

The stability of fixed points in the model can be analyzed from the sign and structure of the drift function $v(c)$, which governs the deterministic evolution of the fraction $c$ of agents in the up-state. The explicit expression is
\begin{equation}
v(c) = (1 - p)\left[(1 - c) c^q - c (1 - c)^q\right] + p(s - c). \label{eq:drift_A}
\end{equation}
The sign of $v(c)$ determines the direction of deterministic flow: $v(c) > 0$ drives the system toward $c = 1$, while $v(c) < 0$ drives it toward $c = 0$. The external field introduces asymmetry into the dynamics, shifting the position and stability of fixed points compared to the unbiased $q$-VM.

We revisit the fixed-point condition $v(c) = 0$. For $q = 1$ and $p > 0$, the drift is linear, and the unique stable fixed point is $c = s$. For $q > 1$, several special cases illustrate the structure of the fixed points:

Case $s = 0$: The system admits a trivial fixed point at $c = 0$. Its stability can be confirmed by evaluating the derivative,
    \begin{align}
        v'(0) = & \lim_{c \to 0} \frac{v(c)}{c} \nonumber \\
        = & \lim_{c \to 0} \Bigl\{(1 - p)(1 - c) \left[c^{q - 1} - (1 - c)^{q - 1}\right] - p\Bigr\}. \nonumber
    \end{align}
    For $q = 1$, $v'(0) = -p$; for $q > 1$, the dominant contribution yields $v'(0) = -1$. In both cases, $v'(0) < 0$, confirming that $c = 0$ is a stable fixed point. A second, nontrivial fixed point satisfies
    \begin{equation}
        (1 - c)^q - (1 - c) c^{q - 1} = -\frac{p}{1 - p},
    \end{equation}
    with location and stability depending on $p$ and $q$.

Case $s = 1$: A symmetric argument yields a fixed point at $c = 1$, which is also stable. Evaluating the derivative:
    \begin{align}
        v'(1) = & \lim_{c \to 1} \frac{v(c)}{c - 1} \nonumber \\
        = & \lim_{c \to 1} \Bigl\{-c(1 - p)\left[c^{q - 1} - (1 - c)^{q - 1}\right] - p\Bigr\}. \nonumber
    \end{align}
    Again, $v'(1) < 0$ for both $q = 1$ and $q > 1$. A second fixed point, if it exists, satisfies
    \begin{equation}
        c^q - c(1 - c)^{q - 1} = -\frac{p}{1 - p}.
    \end{equation}

Case $s=\tfrac{1}{2}$: The fixed point $c = 1/2$ is always a solution due to symmetry. Its stability is determined by
    \begin{align}\label{eq:vprime_half}
        v'(1/2) =  \lim_{c \to \frac{1}{2}} \frac{v(c)}{c - \frac{1}{2}} 
        =  (1 - p)\frac{q - 1}{2^{q - 1}} - p.
    \end{align}
    Setting $v'(1/2) = 0$ yields the critical point
    \begin{equation}\label{eq:p_c}
        p_c = \frac{q - 1}{q - 1 + 2^{q - 1}},
    \end{equation}
    as the same with Eq.~\eqref{eq:pc_result}. For $q = 1$, $v'(1/2) = -p < 0$, so $c = 1/2$ is stable. For $q > 1$, the point becomes unstable if $p < p_c$ and stable if $p > p_c$.

When $p < p_c$, the function $v(c)$ becomes symmetric around $c = 1/2$ and develops two stable fixed points symmetrically located at $c = 1/2 \pm \delta$. To estimate $\delta$, we expand $v(c)$ near $c = 1/2$:
\begin{equation}\label{eq:delta_expand}
    v(\tfrac{1}{2}+ \delta) = A \delta - B_{2k+1}(q) \delta^{2k + 1} + \mathcal{O}(\delta^{2k + 3}),
\end{equation}
where even-order terms vanish due to symmetry. Here, $A$ is the linear coefficient, and $B_{2k+1}(q)$ is the first nonvanishing odd-order coefficient, with $k \geq 1$. The nontrivial fixed point is then given by
\begin{equation}
    \delta \approx \left(\frac{A}{B_{2k+1}(q)}\right)^{\frac{1}{2k}},
\end{equation}
with $A = (1 - p)(q - 1)2^{1 - q} - p$ for all values of $q > 1$. For illustration, we note that for $q = 2, 3$, the leading nonlinear coefficient is $B = 4(1 - p)(q - 1)2^{1 - q}$.

Let $F(c;p,s,q)\equiv v(c)$ be the mean–field drift. Saddle–node (fold) loci are obtained from the degeneracy conditions
$F(c;p,s,q)=0$ and $\partial F/\partial c=0$.
Writing $G(c,q)=c^{q}(1-c)-(1-c)^{q}c$ and $G'=\partial G/\partial c$, a convenient parameterization of the folds is
\begin{equation}
  p(c)=\frac{G'(c,q)}{1+G'(c,q)},\qquad
  s(c)=c-\frac{G(c,q)}{G'(c,q)}, \label{eq:fold-param}
\end{equation}
with $c\in(0,1)$, $G'(c,q)\neq0$, and $(p,s)$ restricted to $(0,1)\times(0,1)$.
These curves bound the multistable region in the $(s,p)$ plane: a cusp for $1<q<5$ (supercritical regime) and a butterfly structure for $q>5$ (subcritical regime). At $s=\tfrac12$ the symmetric branch $c=\tfrac12$ must be treated explicitly (Eq.~\eqref{eq:stat} reduces to $0/0$ there); its stability along $p$ follows from the sign of $v'(c)$.

\begin{figure}[tb]
    \centering
    \includegraphics[width=\linewidth]{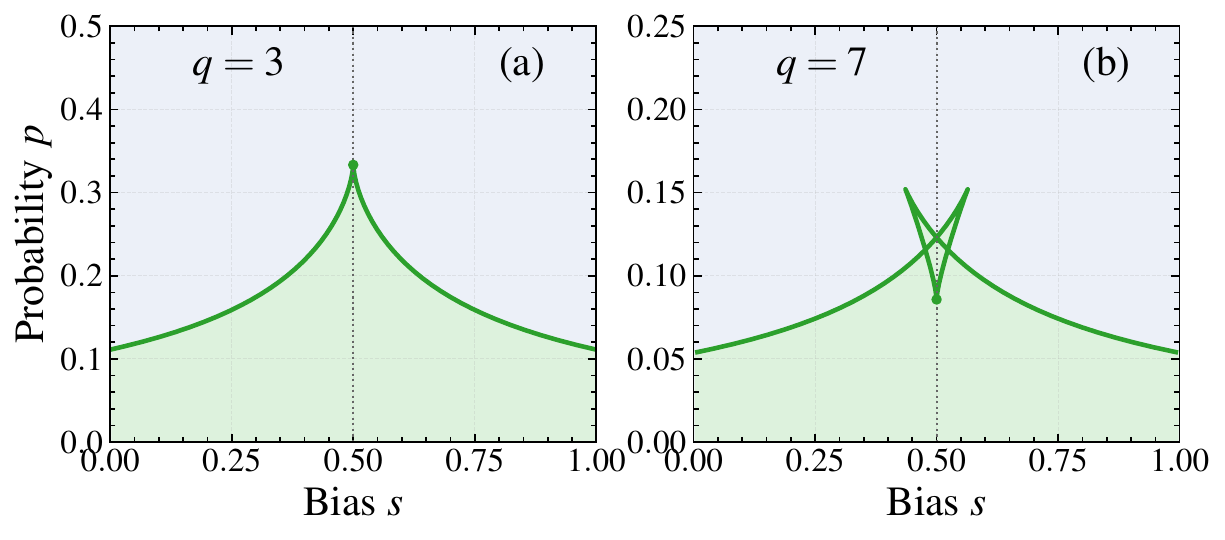}
    \caption{Fold (saddle–node) loci in the $(s,p)$ plane from Eq.~\eqref{eq:fold-param}.
    Green curves are fold lines; the dotted vertical marks $s=\tfrac12$; the dot at $s=\tfrac12$ is $p_c(q)$.
    Light–green shading indicates multistability (three fixed points: two stable + one unstable); the bluish region is monostable.
    (a) $q=3$ (supercritical): a cusp with tip at $p_c$, shrinking bistable interval as $\delta s$ increases.
    (b) $q=7$ (subcritical): a butterfly–like structure near $s=\tfrac12$; the outer folds bound a multistable region and the inner ``X" signals additional folds (a small pocket with five fixed points may appear very close to $s=\tfrac12$ for low $p$).}
    \label{fig:phase_ps}
\end{figure}

\begin{figure}[tb]
    \centering
    \includegraphics[width=\linewidth]{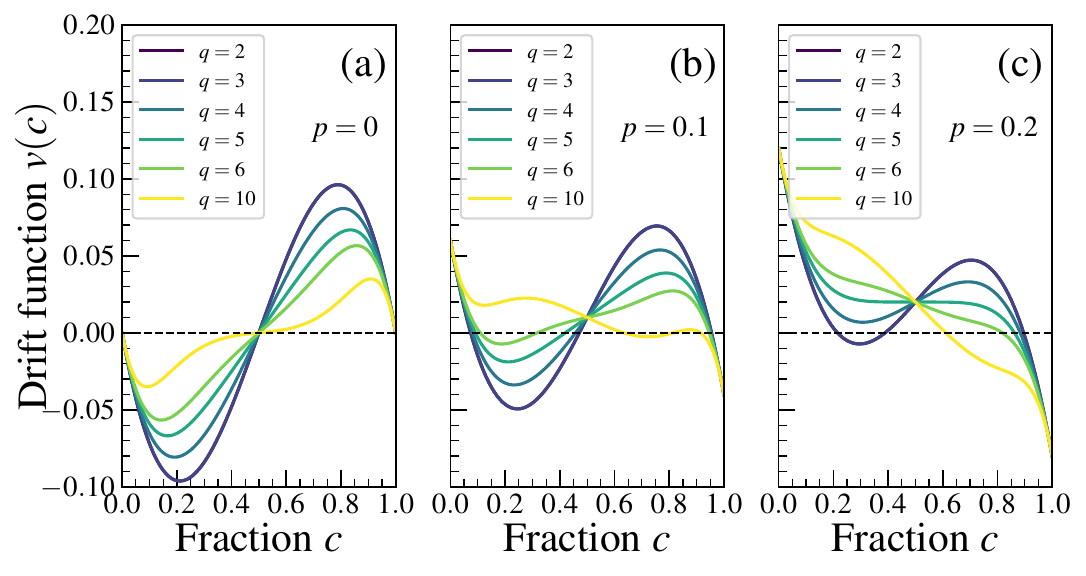}
    \caption{Drift function $v(c)$ for $s=0.60$ and several values of $q$. Each panel uses a different independence probability $p$. (a) $p=0$ (field inactive): $v(c)$ is antisymmetric with stable fixed points at $c=0$ and $c=1$ and an unstable one at $c=\tfrac12$ (independent of $s$). (b) $p=0.1$: the bias $s>\tfrac12$ breaks the symmetry and shifts the zeros; depending on $q$, one observes either a single fixed point or three (two stable separated by one unstable), i.e., bistability. (c) $p=0.2$: the field term is stronger and, for moderate/large $q$, only one stable fixed point remains, located near the bias (the zero $c^*$ with $v(c^*)=0$ satisfies $c^*\gtrsim s$); for small $q$ (e.g., $q=2,3$) a three–fixed–point structure may still persist.}
    \label{fig:drift_shape}
\end{figure}

To interpret the fold boundaries, we inspect one-dimensional slices of the dynamics at fixed bias. Figure~\ref{fig:drift_shape} plots the drift $v(c)$ at $s=0.60$ for several $q$, with panels (a)–(c) corresponding to $p=0$, $0.1$, and $0.2$, respectively. A stationary state $c^{*}$ satisfies $v(c^{*})=0$ (stable if $v'(c^{*})<0$). As $p$ increases at fixed $s$, zeros of $v(c)$ are created or annihilated precisely when the point $(s,p)$ crosses the fold curves in Fig.~\ref{fig:phase_ps}. For $p=0$ the field contribution vanishes and $v(c)$ is antisymmetric. For $q>1$ there are stable fixed points at $c=0$ and $c=1$ and an unstable one at $c=\tfrac12$. At $p=0.1$ the bias $s>\tfrac12$ breaks the up–down symmetry and shifts the zeros; depending on $q$, one observes either a single fixed point or three (two stable separated by one unstable), which implies bistability. At $p=0.2$ the field term is stronger and, for moderate to large $q$, only one stable fixed point remains near the bias (the zero $c^{*}$ with $v(c^{*})=0$ satisfies $c^{*}\gtrsim s$); for small $q$ (for example, $q=2$ or $3$) the three–fixed–point structure may still persist.

\section{\label{app:app_mot}Consensus Time}

In this Appendix, we derive an analytical expression for the consensus time $T$ under the deterministic, large-$N$ approximation. In this limit, the diffusion term $D(c)\sim1/N$ in the backward Kolmogorov equation is negligible compared to the drift term, exposing the dependence of $T$ on model parameters.

The consensus time $T(c)$ is the mean first‐passage time to either absorbing state $c = 0$ or $c = 1$. For $s = 0$ the system deterministically orders to $c = 0$, whereas for $s = 1$ it orders to $ c = 1$. Each update involves a single agent, changing the opinion fraction by $\delta c=1/N$ and advancing time by $\delta t=1/N$. The backward recursion for $T(c)$ is
\begin{align}
T(c) ={}& R(c)\bigl[T(c+\delta c)+\delta t\bigr]
         +L(c)\bigl[T(c-\delta c)+\delta t\bigr]\nonumber\\
        & +\bigl[1 - R(c) - L(c)\bigr]\bigl[T(c)+\delta t\bigr],
\end{align}
where $R(c)$ and $L(c)$ are the transition probabilities.

Expanding $T(c\pm\delta c)$ to second order in $\delta c$ and collecting terms leads to the backward Kolmogorov equation in the continuum limit:
\begin{equation} \label{eq:relax_rel_app}
    v(c)\, T'(c) + D(c)\, T''(c) + 1 = 0,
\end{equation}
with boundary conditions $T(0) = T(1) = 0$. Here, $v(c) \equiv R(c) - L(c)$ and $D(c) \equiv [R(c) + L(c)]/(2N)$ are the drift and diffusion functions, respectively.

Neglecting the diffusion term yields a first-order differential equation, which integrates to
\begin{equation}\label{eq:relax_time_app}
T(c(0)) \approx \int_{c(0)}^{1-1/N} \frac{dc}{v(c)},
\end{equation}
where the upper limit implements the finite-size regularization at the absorbing boundary. For $s=1$, $c=1$ is absorbing; since $c$ takes values on the lattice $\{0,1/N,\ldots,1\}$, the nearest interior point is $1-1/N$, which we adopt as the cutoff.

For $s = 1$, the drift simplifies to
\begin{equation}
    v(c) = \left(1 - c\right)\left[\left(1 - p\right)\left(c^q - c\left(1 - c\right)^{q - 1}\right) + p\right], \label{eq:drift_s1_app}
\end{equation}
and  for $q = 1$, the nonlinear term vanishes, and $v(c) = p\left(1 - c\right)$. Substituting into Eq.~\eqref{eq:relax_time_app} gives
\begin{equation}\label{eq:exit_time_linear_app}
    T(N, p) \approx \dfrac{1}{p} \int_{c(0)}^{1-1/N}  \dfrac{du}{1-u} \sim \dfrac{1}{p} \ln N,
\end{equation}
which corresponds to Eq.~\eqref{eq:relax_linear} in the main text. For $q > 1$, the integral in Eq.~\eqref{eq:relax_time_app} can be evaluated by changing variables $u = 1 - c$, yielding
\begin{align}
    T(N, p) &\approx \int_{1/N}^{u(0)} \frac{du}{u F(1 - u)},
\end{align}
with
\begin{align*}
    F(c) &= (1 - p)\left[c^q - c\left(1 - c\right)^{q - 1}\right] + p.
\end{align*}
Using partial fraction decomposition:
\begin{equation}
    F(1 - u) = \prod_{i = 1}^{q} \left(1 - r_i u\right)^{\mathcal{A}_i},
\end{equation}
the integral becomes
\begin{align}
    T(N, q, p) &\approx \int_{1/N}^{u(0)} \frac{du}{u} \prod_{i = 1}^{q} \left(1 - r_i u\right)^{-\mathcal{A}_i}  \nonumber \\
    &= \int_{1/N}^{u(0)} \dfrac{du}{u}\nonumber + \sum_{i}^{q} \dfrac{\mathcal{A}_i}{r_i} \int_{1/N}^{u} \dfrac{du}{1-r_iu}\\
    &= \ln N - \sum_{i = 1}^{q} \frac{\mathcal{A}_i}{r_i^2} \ln\left[1 - r_i \left(1 - c(0)\right)\right].
\end{align}
Thus, the consensus time admits the compact form:
\begin{equation}
    T(N, c(0), q, p) \approx \ln N + \mathcal{C}(c(0), q, p),
\end{equation}
where
\begin{equation}
    \mathcal{C}(c(0), q, p) = - \sum_{i = 1}^{q} \frac{\mathcal{A}_i}{r_i^2} \ln\left[1 - r_i \left(1 - c(0)\right)\right].
\end{equation}
This expression corresponds to Eq.~\eqref{eq:relax_nonlinear} in the main text.

Although the coefficients $\mathcal{A}_i$ and $r_i$ are difficult to obtain analytically for general $q$, we illustrate the result for $q = 2$ and $q = 3$, where
\begin{equation}
    F(1 - u) = 1 - 3(1 - p)u + 2(1 - p)u^2 = (1 - r_1 u)(1 - r_2 u),
\end{equation}
with
\begin{equation}
    r_{1,2} = \frac{3(1 - p) \pm \sqrt{(1 - p)(1 - 9p)}}{2}.
\end{equation}
Substituting into the general formula for $\mathcal{C}(c(0), p)$ yields:
\begin{align}
\mathcal{C}(c(0), p) = & \frac{3(1 - p)r_1 - 2(1 - p)}{r_1(r_1 - r_2)} \ln\left[1 - r_1\left(1 - c(0)\right)\right] \nonumber \\
& + \frac{2(1 - p) - 3(1 - p)r_2}{r_2\left(r_1 - r_2\right)} \ln\left[1 - r_2\left(1 - c(0)\right)\right].
\end{align}
For the balance case $c(0) = 1/2$, this further simplifies to:
\begin{align}
    \mathcal{C}(p) = - \frac{3}{2} \sqrt{\frac{1 - p}{1 - 9p}} \ln \left(\frac{1 + 3p - D}{1 + 3p + D}\right) - \frac{1}{2} \ln p,
\end{align}
where $D = \sqrt{(1 - p)(1 - 9p)}$, and valid for $0 < p < 1/9$.

\section{Consensus Time for Linear Model and Weak-Selection Limit}

In the case of the linear VM, the drift and diffusion coefficients simplify to
\begin{equation} \label{eq:drif_diff_q1_app}
\begin{aligned}
v(c) & = p\left(1 - c\right), \\
D(c) & = \frac{(1 - c)\left[2c(1 - p) + p\right]}{2N}.
\end{aligned}
\end{equation}
Substituting these expressions into the backward Kolmogorov equation yields
\begin{equation}\label{eq:T_c_app}
(1 - c) \left[ p\,T'(c) + \frac{1}{2N} \left(2c(1 - p) + p \right) T''(c) \right] + 1 = 0.
\end{equation}

In the weak-selection limit ($p \ll 1$), $T(c)$ can be expanded perturbatively in powers of $p$:
\begin{equation}\label{eq:T_exp_app}
T(c) = T_0(c) + p\,T_1(c) + \mathcal{O}(p^2),
\end{equation}
where $T_0(c(0))$ is the consensus time for the classical VM:
\begin{equation}\label{eq:T0_app}
T_0(c(0)) = -N \left[ c(0) \ln c(0) + \left(1 - c(0)\right) \ln \left(1 - c(0) \right) \right].
\end{equation}
Substituting Eqs.~\eqref{eq:T_exp_app} and \eqref{eq:T0_app} into Eq.~\eqref{eq:T_c_app}, and expanding to first order in $p$, yields
\begin{align}\label{eq:T1_eq}
(1 - c) \Bigg[ T_0'(c) + \frac{1 - 2c}{2N} T_0''(c) + \frac{c}{N} T_1''(c) \Bigg] = 0.
\end{align}
Evaluating derivatives of $T_0(c)$ and substituting into Eq.~\eqref{eq:T1_eq}, we obtain
\begin{equation}
T_1''(c) = \frac{N^2}{c} \ln\left(\frac{c}{1 - c}\right) + \frac{N\left(1 - 2\,c\right)}{2\,c^2\left(1 - c\right)}.
\end{equation}
Integrating twice and applying appropriate boundary conditions yields the first-order correction:
\begin{equation}
T_1(c(0)) = -N^2 \left[ \mathrm{Li}_2\left(1 - c(0)\right) - \mathrm{Li}_2(c(0)) + \frac{\pi^2}{3}c(0) - \frac{\pi^2}{6} \right],
\end{equation}
where $\mathrm{Li}_2(\cdot)$ is the dilogarithm or Spence’s function. Thus, the asymptotic expression for the consensus time up to the first-order correction is:
\begin{align}\label{eq:T1_final_app}
T(c(0)) \approx & -N \left[ c(0) \ln c(0) + \left(1 - c(0)\right) \ln \left(1 - c(0)\right) \right] \nonumber \\
& - p N^2 \left[ \mathrm{Li}_2\left(1 - c(0)\right) - \mathrm{Li}_2(c(0)) + \frac{\pi^2}{3}c(0) - \frac{\pi^2}{6} \right].
\end{align}
Analogously, for $s = 0$, the consensus time up to first-order correction reads:
\begin{align}\label{eq:T1_final_app_s0}
T(c(0)) \approx & -N \left[ c(0) \ln c(0) + \left(1 - c(0)\right) \ln \left(1 - c(0)\right) \right] \nonumber \\
& + p N^2 \left[ \mathrm{Li}_2\left(1 - c(0)\right) - \mathrm{Li}_2(c(0)) + \frac{\pi^2}{3}c(0) - \frac{\pi^2}{6} \right].
\end{align}

\begin{figure}[tb]
    \centering
    \includegraphics[width=0.62\linewidth]{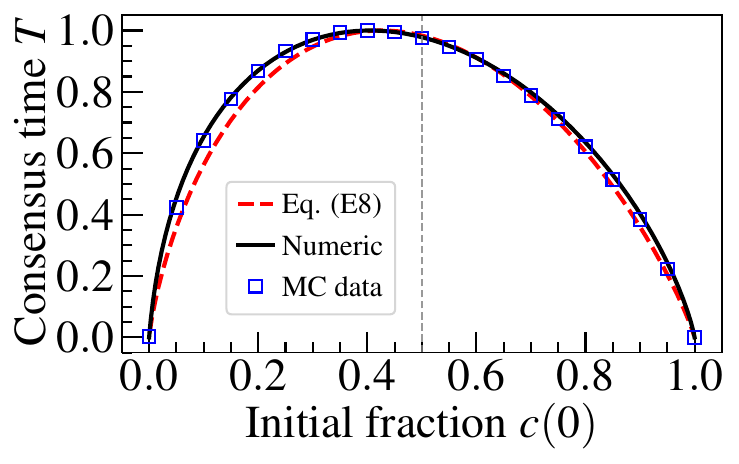}
    \caption{Comparison of the analytical approximation for the normalized consensus time $T$ from Eq.~\eqref{eq:T1_final_app} with the numerical solution and MC simulations, for $N = 50$ and $p = 0.01$.}
    \label{fig:mot_p}
\end{figure}

These expressions characterize the neutral dynamics of the classic VM and the leading-order correction due to a weak external bias. A comparison between the analytical approximation given by Eq.~\eqref{eq:T1_final_app}, the numerical solution of the full differential equation, and MC simulations is exhibited in Fig.~\ref{fig:mot_p}.

\section{\label{appendix_F} Disordering Time}

To derive the disordering time of the model, we begin by considering the drift function at the symmetric point $s=\tfrac{1}{2}$. In this case, the drift takes the form
\begin{equation}
v(c) = (1 - p) \left[ (1 - c) c^q - c (1 - c)^q \right] - p\left(c - \tfrac{1}{2}\right)
\end{equation}
The function $v(c)$ has a simple root at $c = 1/2$, which corresponds to a stable fixed point for all $p > p_c$ when $q > 1$, and for all $p > 0$ when $q = 1$.

In the vicinity of this balanced configuration, the drift function can be expanded as
\begin{equation}
\left.\frac{v(c)}{2c - 1} \right|_{\lim c \to 1/2} = (1 - p)(q - 1)2^{-q} - \frac{p}{2}.
\end{equation}
Substituting this into Eq.~\eqref{eq:relax_time_app} and evaluating the integral from $c = 1$ to $c = 1/2 + 1/\sqrt{N}$ yields
\begin{align}\label{eq:pola_general}
T(q, p, N) &\approx \frac{1}{(1 - p)(q - 1)2^{-q} - \frac{p}{2}} \int_{1}^{1/2 + 1/\sqrt{N}} \frac{dc}{2c - 1} \nonumber \\
&\sim  \frac{\ln N}{2p - (1 - p)(q - 1) 2^{2 - q}},
\end{align}
which corresponds to Eq.~\eqref{eq:dis_voter} in the main text. Furthermore, Eq.~\eqref{eq:pola_general} can be expressed in power-law form as
\begin{align}\label{eq:exit_time_pl_app}
T(q, p, N) & = \frac{\ln N}{2p - (1 - p)(q - 1) 2^{2 - q}} \nonumber \\
&= \frac{\ln N}{(p - p_c)\left[2 + (q - 1) 2^{2 - q}\right]} \nonumber \\
& \sim |p - p_c|^{-1} \ln N,
\end{align}
where the critical probability $p_c$ is given by Eq.~\eqref{eq:p_c}. This Eq.~\eqref{eq:exit_time_pl_app} corresponds to Eq.~\eqref{eq:exit_time_pl} in the main text.

\begin{figure}[tb]
    \centering
    \includegraphics[width=0.63\linewidth]{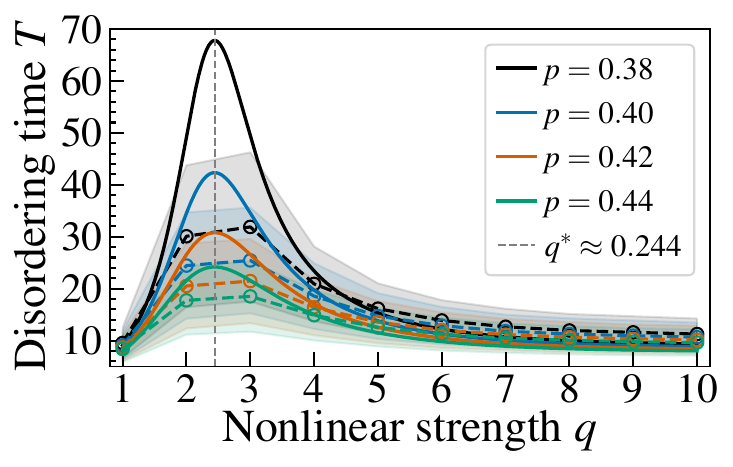}
    \caption{Disordering time $T$ as a function of the nonlinear strength $q$ for independence probabilities $p>p_c$, computed for a system of size $N=10^3$ and averaged over $10^4$ independent realizations. Solid lines show analytical predictions from Eq.~\eqref{eq:pola_general}, markers connected by dashed lines correspond to MC results, and shaded regions indicate one standard deviation. All datasets exhibit a peak at $q^*\approx2.44$.}
    \label{fig:polar_var_p}
\end{figure}

We now analyze the behavior of the disordering time $T$ as a function of the nonlinearity strength $q$, under two distinct scenarios: (i) fixed $p > p_c$, and (ii) tuned $p(q)$ such that the distance from criticality, $|p - p_c(q)|$, remains constant across different values of $q$. In the first scenario, the value $| p-p_c |$ increases with $q$, since the critical value $p_c$ decreases monotonically with $q$. Differentiating Eq.~\eqref{eq:pola_general} with respect to $q$ yields
\begin{equation}
T'(N,q,p) = -\frac{\left(1 - p\right)2^{2 - q} \left[ 1 - \left(q - 1\right)\ln 2 \right]}{\left[2p - \left(1 - p\right)\left(q - 1\right)2^{2 - q} \right]^2} \ln N.
\end{equation}
Setting $T'(N,q,\alpha) = 0$, the numerator vanishes when
\begin{equation}
1 - (q - 1)\ln 2 = 0 \quad \implies \quad q^{*} = 1 + \frac{1}{\ln 2} \approx 2.44.
\end{equation}
The value $q^*$ corresponds to a local maximum of the disordering time $T(N, q, p)$ for all values of $p > p_c(q)$, as illustrated in Fig.~\ref{fig:polar_var_p}. Accordingly, the peak disordering time $T_{\text{max}}$ is
\begin{equation}
    T_{\text{max}} \sim \dfrac{e \ln 2}{2 \left[ \left(e \ln 2 + 1 \right) p - 1\right]} \ln N.
\end{equation}

In the second scenario, we fix the distance from criticality across all $q$ by tuning $p$ such that $p(q) = \alpha \, p_c(q)$ with $\alpha > 1$. Substituting this expression into Eq.~\eqref{eq:pola_general} yields
\begin{align}\label{eq:T_tunnes_app}
T(N, q, \alpha) & \sim \frac{\ln N}{2\alpha p_c - (1 - \alpha p_c)(q - 1)2^{2 - q}} \nonumber \\
&= \frac{2^{q - 2}}{(\alpha - 1)(q - 1)} \ln N,
\end{align}
which corresponds to Eq.~\eqref{eq:T_tunned} in the main text.

Taking the derivative of $T(N, q, \alpha)$ with respect to $q$, we find
\begin{equation}
T'(N,q,\alpha) \sim \frac{2^{q - 2} \left[(q - 1)\ln 2 - 1\right]}{(\alpha - 1)(q - 1)^2} \ln N ,
\end{equation}
which vanishes when
\begin{equation}
(q - 1)\ln 2 - 1 = 0 \quad \implies \quad q^{*} = 1 + \frac{1}{\ln 2} \approx 2.44.
\end{equation}
Thus, $q^{*}$ also represents the unique minimum of $T(N, q, \alpha)$ for all $\alpha > 1$. Hence, the minimum disordering time scales as
\begin{equation}
    T_{\text{min}} \sim \dfrac{e \ln 2}{2 \left(\alpha-1\right)} \ln N.
\end{equation}

\section{\label{app:app_exip}Exit Probability}

The backward Kolmogorov equation for the exit probability $E(c)$ follows directly from the discrete recursion
\begin{equation}
E(c) = R(c)E(c + \delta c)+L(c)E(c - \delta c)+\bigl[1 - R(c) - L(c)\bigr]E(c).
\end{equation}
Expanding $E(c\pm\delta c)$ to second order in $\delta c$ and passing to the continuous limit $\delta c\to0$ yields the ordinary differential equation
\begin{equation}\label{eq:exit_kol}
v(c) \dfrac{dE(c)}{dc}+D(c) \dfrac{d^2E(c)}{dc^2}=0,
\end{equation}
subject to the boundary conditions $E(0)=0,E(1)=1.$ Equivalently, one may write $E(c(0))$ in integral form as
\begin{equation}\label{eq:exit_integral}
E(c(0)) =
\frac{\displaystyle\int_{0}^{c(0)}\exp\Bigl[-\int_{0}^{y}\frac{v(u)}{D(u)}du\Bigr]dy}
{\displaystyle\int_{0}^{1}\exp\Bigl[-\int_{0}^{y}\frac{v(u)}{D(u)}du\Bigr]dy}.
\end{equation}

\subsection{Case for linear model, $q  = 1$}
For $q = 1$, an exact expression for the exit probability can be obtained from Eq.~\eqref{eq:exit_integral}, as the drift and diffusion functions admit a closed-form expression given by Eq.~\eqref{eq:drif_diff_q1_app}. Defining
\begin{align}
    K(u)   = \int \dfrac{v(u)}{D(u)} du = \dfrac{Np}{(1-p)} \ln \left[ 2(1-p)u + p \right],
\end{align}
the integral form of the exit probability becomes
\begin{align}
    E(c(0)) & =   \dfrac{ \displaystyle \int_{0}^{c(0)} \left[ 2\left(1-p\right)c + p \right]^{-\frac{Np}{1-p}} \, dc}{\displaystyle\int_{0}^{1} \left[ 2\left(1-p\right)c + p \right]^{-\frac{Np}{1-p}} \, dc} \nonumber \\
    &= \dfrac{\displaystyle\int_{p}^{2\left(1-p\right)c(0)+p}y^{-\frac{Np}{1-p}}dy}{\displaystyle\int_{p}^{2-p}y^{-\frac{Np}{1-p}} dy}  \nonumber \\ & = 
\dfrac{\left[2\left(1-p\right)c(0) + p \right]^{1 - \frac{Np}{1-p}} - p^{1 - \frac{Np}{1-p}}}{\left(2-p \right)^{1 - \frac{Np}{1-p}} - p^{1 - \frac{Np}{1-p}}}.
    \label{eq:exit_s1}
\end{align}
For the second case, where $s = 0$, the function $K(u)$ becomes
\begin{equation}
    K(u) = \dfrac{Np}{(1-p)} \ln \left[ (2-p) - 2u (1-p) \right],
\end{equation}
which leads to the exit probability
\begin{equation}\label{eq:exit_s0}
    E(c(0)) = \dfrac{\left[\left(2 - p\right) - 2c(0)\left(1- p\right)\right]^{1 - \frac{Np}{1-p}} - \left(2 - p\right)^{1 - \frac{Np}{1-p}}}{p^{1 - \frac{Np}{1-p}} - \left(2 - p\right)^{1 - \frac{Np}{1-p}}}.
\end{equation}
Equations~\eqref{eq:exit_s1} and \eqref{eq:exit_s0} correspond to the general expressions provided in Eq.~\eqref{eq:exit_p_ori} of the main text. In the limit $p \to 0$, both expressions reduce to the well-known result for the original VM, namely $E(c(0)) = c(0)$.

\subsection{Case for $q > 1$ and $p = 0$}
For $q > 1$, one can derive an approximate analytical expression for the exit probability by applying the saddle‐point (Laplace) approximation to Eq.~\eqref{eq:exit_integral}. In the limit $p = 0$, the ratio of drift to diffusion simplifies to
\begin{equation}
    \dfrac{v(c)}{D(c)} =  2N \frac{c^{q-1} - (1 - c)^{q-1}}{c^{q-1} + (1 - c)^{q-1}}.
\end{equation} 
Defining
\begin{align}
    \psi (r) = & \frac{r^{q-1} - (1 - r)^{q-1}}{r^{q-1} + (1 - r)^{q-1}},
\end{align}
and
\begin{equation}
    K(r) = 2N
\int_{1/2}^{r}\psi(u)\,\mathrm{d}u\,.
\end{equation}
Because $\psi(r)$ is antisymmetric about $r=\tfrac12$ ($\psi(1/2)=0$, $\psi(r)<0$ for $r<1/2$, and $\psi(r)>0$ for $r>1/2$), $K(r)$ attains its unique global minimum at $r=1/2$.

Since the exit probability involves an integral weighted by $\exp[-K(r)]$, its main contribution originates from the vicinity of the minimum of $K(r)$, where the integrand is maximal.  Hence, one may apply the saddle‐point approximation to evaluate the integral.

Define
\begin{equation}
r = \tfrac12 + u\,,\qquad |u|\ll1.    
\end{equation}
Expanding $(\tfrac12\pm u)^{q-1}$ to first order in $u$ gives
\begin{equation}
\Bigl(\tfrac12 \pm u\Bigr)^{q-1}
\approx
2^{1-q}\bigl[1 \pm 2(q-1)u\bigr].
\end{equation}
Substituting into the numerator and denominator of $\psi(r)$ yields
\begin{align}
r^{q-1} - (1-r)^{q-1}
&\approx (q-1)2^{3-q}u\\
r^{q-1} + (1-r)^{q-1}
&\approx 2^{2-q}.
\end{align}
Accordingly,
\begin{align}
\psi(r)
= \frac{r^{q-1} - (1-r)^{q-1}}{r^{q-1} + (1-r)^{q-1}}
& \approx\frac{(q-1)2^{3-q}u}{2^{2-q}}
\nonumber \\
& = (q-1)(2r - 1),
\end{align}
where we have used $2r-1=2u$.  This linearized form of $\psi(r)$ around $r=\tfrac12$ provides the Gaussian integral kernel for the saddle‐point evaluation of the exit‐probability integral.
 
We now turn to the quadratic expansion of $K(r)$ around its minimum at $r=\tfrac12$.  From the definition $K'(r)=2N\,\psi(r)$ and the linearized form of $\psi(r)$, one immediately obtains
$K'(r)=2N\psi(r), \,K'\bigl(\tfrac12\bigr)=0,
\,K''(r)=2N\psi'(r), \,K''\bigl(\tfrac12\bigr)=4N(q-1)$.
Hence, a second‐order Taylor expansion of $K(r)$ about $r=\tfrac12$ gives
\begin{align}
   K(r) & \approx K\bigl(\tfrac12\bigr)+\tfrac12\,K''\bigl(\tfrac12\bigr)(r-\tfrac12)^{2} \nonumber \\
&=K\bigl(\tfrac12\bigr)+2N(q-1)(r-\tfrac12)^{2}. 
\end{align}
Since the constant term $K(\tfrac12)$ cancels between numerator and denominator in the exit‐probability ratio, the integrand in Eq.~\eqref{eq:exit_integral} reduces to
\begin{equation}
 \exp\bigl[-K(r)\bigr]\;\propto\;\exp\bigl[-2N\,(q-1)\,(r-\tfrac12)^{2}\bigr].   
\end{equation}
Setting $r=\tfrac12+u$ and, in the limit $N\gg1$, extending the integration limits to $\pm\infty$, we approximate
\begin{equation}
\exp\bigl[-K(r)\bigr]\;\approx\;\exp\bigl[-2N\,(q-1)\,u^{2}\bigr].   
\end{equation}
Finally, for an initial condition $c(0)=\tfrac12+\Delta$ with $\Delta\ll1$, Eq.~\eqref{eq:exit_integral} becomes a ratio of Gaussian integrals:
\begin{align}\label{eq:exit_general}
E(c(0))
&\approx
\frac{\displaystyle\int_{-\infty}^{\Delta}\exp\left[-2N\left(q-1\right)u^{2}\right]du}
{\displaystyle\int_{-\infty}^{\infty}\exp\left[-2N(q-1)u^{2}\right]du}
\nonumber\\
&=
\dfrac{\tfrac12\sqrt{\frac{\pi}{2N(q-1)}}\Bigl[1+\mathrm{erf}\bigl(\sqrt{2N(q-1)}\Delta\bigr)\Bigr]}
{\sqrt{\frac{\pi}{2N(q-1)}}}
\nonumber\\
&=\tfrac12\bigl[1+\mathrm{erf}\bigl(\sqrt{2N_{\rm eff}\left(q-1\right)}\left(c(0)-\tfrac12\right)\bigr)\bigr],
\end{align}
which recovers the form presented in Eq.~\eqref{eq:exit_general_eff} of the main text.

It follows immediately from Eq.~\eqref{eq:exit_general} that the width of the crossover region in $E(c(0))$ around $c(0)=\tfrac12$ scales as $\left[N(q-1)\right]^{-1/2}$, and hence vanishes in the thermodynamic limit.  Indeed, as $N\to\infty$, the error‐function profile sharpens into a Heaviside step,
\begin{equation}
 E(c(0))\;\longrightarrow\;\Theta\bigl(c(0)-\tfrac12\bigr),   
\end{equation}
so that any infinitesimal bias $\Delta=c(0)-\tfrac12$ suffices to select one of the two absorbing states deterministically. This emergent ``all‐or‐nothing" behavior reflects the dominance of nonlinear drift for $q>1$, and we emphasize that the saddle‐point approximation leading to Eq.~\eqref{eq:exit_general} holds only in this regime of genuine nonlinearity, where the disordering transition is discontinuously sharp.

\subsection{Case for $q>1$ and $p\neq0$}
For nonzero $p$, the interior stationary point shifts to $r^\ast(p,s,q)$ and the local curvature can be reduced so that the integrand may receive significant weight from a broader neighborhood. In these regimes, the quadratic Laplace approximation is not uniformly accurate and can break down, especially near boundaries or fold loci where stationary points coalesce. Consequently, a simple closed form is not generally available for this case. We therefore evaluate Eq.~\eqref{eq:exit_integral} by direct numerical quadrature. Uniform asymptotics, e.g., boundary-layer matching or Airy-type approximations, could be developed but lie beyond our scope.

\nocite{*}

\bibliography{apssamp}% Produces the bibliography via BibTeX.

\end{document}